\documentclass{aa}
\usepackage{natbib}
\usepackage{graphicx}
\bibpunct{(}{)}{;}{a}{}{,} % to follow the A&A style

\begin{document}

\title{Where are the hot ion lines in classical T Tauri stars formed~?}

\author{H.M. G\"unther \and J.H.M.M. Schmitt}  %\inst{1} 
\offprints{H. M. G\"unther,\\ \email{moritz.guenther@hs.uni-hamburg.de}}
\institute{Hamburger Sternwarte, Universit\"at Hamburg, Gojenbergsweg 112, 21029 Hamburg, Germany}
\date{Received 14 September 2007 / accepted 15 January 2008}
\abstract{Classical T Tauri stars (hereafter CTTS) show a plethora of in- and outflow signatures in a variety of wavelength bands.}{In order to constrain gas velocities and temperatures, we analyse the emission in the hot ion lines.}{We use all available archival \emph{FUSE} spectra of CTTS to measure the widths, fluxes and shifts of the detected hot ion lines and complement these data with \emph{HST/GHRS} and \emph{HST/STIS} data. We present theoretical estimates of the temperatures reached in possible emission models such as jets, winds, disks and accretion funnels and look for correlations with X-ray lines and absorption properties.}{We find line shifts in the range from -170~km~s$^{-1}$ to +100~km~s$^{-1}$. Most linewidths exceed the stellar rotational broadening. Those CTTS with blue-shifted lines also show excess absorption in X-rays. CTTS can be distinguished from main sequence (hereafter MS) stars by their large ratio of the \ion{O}{vii} to \ion{O}{vi} luminosities.}{No single emission mechanism can be found for all objects. The properties of those stars with blue-shifted lines are compatible with an origin in a shock-heated dust-depleted outflow.}
\keywords{stars: formation -- stars: winds/outflows -- Ultraviolet: stars}
\titlerunning{Formation of hot ion lines in CTTS}
\maketitle

\section{Introduction}
\label{introduction}

T Tauri stars (TTS) are young ($<10\;\mathrm{Myr}$), low mass ($M_*<3M_{\sun}$), pre-main sequence stars exhibiting strong H$\alpha$ emission.  The subclass of Classical TTS (CTTS) are accreting material from a surrounding disk.  This disk is 
truncated close to the corotation radius by interaction with the magnetic field. At the inner rim, the material is ionised by the stellar radiation and loaded onto the field lines \citep{1984PASJ...36..105U,1991ApJ...370L..39K} and accelerated to nearly free-fall velocity along an accretion funnel before hitting the stellar surface where an accretion shock forms. Angular momentum may be transferred from the star to the disk by magnetic torque \citep{1994ApJ...429..781S}, at the same time this process is expected to drive outflows possibly in the form of a disk wind or bipolar jets. \citet{2004ApJ...610..920R} performed full magneto-hydrodynamic (MHD) calculations of the accretion geometry, while the accretion shock and its post-shock cooling zone were simulated by different authors \citep{calvetgullbring,lamzin, acc_model}. 

Observational support for the scenario described above is found in data from different wavelength bands. The shocked material is heated up to temperatures in the MK range and should produce copious amounts of X-ray emission.  Density-sensitive X-ray line ratios strongly suggest
a post-shock origin \citep{acc_model} of the cool parts of the plasma observed at X-ray wavelengths, and the cooling matter veils absorption lines in the optical and infrared \citep{calvetgullbring}. 
The geometry of the pre-shock accretion funnel can be traced by H$\alpha$ line profile analysis \citep{2000ApJ...535L..47M}. 
Undoubtedly CTTS also show outflows \citep[a review is given by][]{2007prpl.conf..215B}, however, the precise origin and the physical driving mechanism(s) are
uncertain.  Theoretical models propose a variety of collimated stellar outflows and disk winds to remove angular momentum from the system \citep{1994ApJ...429..781S,2002ApJ...574..232M,2005MNRAS.362..569G,2007prpl.conf..277P}. The outflows can take the form of winds observed from radio to the UV \citep[e.g.][]{2000AJ....119.1881A,2001ApJ...551.1037B,2004AstL...30..413L,2005ApJ...625L.131D,2006ApJ...646..319E}, or collimated, often bipolar jets, again observed at different wavelength bands \citep{1995RMxAC...1....1R,2004ApJ...604..758C,2005ApJ...626L..53G}. 
The dynamics of the surrounding gas have been successfully probed by UV spectroscopy with \emph{HST/GHRS}, \emph{HST/STIS} and \emph{FUSE}, especially in the H$_2$ lines, which show outflows and fluorescence on the hot disk surface \citep{2002ApJ...566.1100A,2002ApJ...567.1013A, 2002ApJ...572..310H, 2005AJ....129.2777H, 2006ApJS..165..256H}. 
In the specific case of TW Hya, \citet{2005ApJ...625L.131D} claim the existence of a warm/hot wind. Their conclusion assumes that the CIII and OVI lines have an intrinsic Gaussian shape, with centroid matching the stellar rest-frame. The observed asymmetry of the CIII and OVI line-profiles can then be explained by the presence of spherically-symmetric and smoothly-accelerated hot wind absorbing the radiation in the blue wing of these particular lines. 
However, \citet{2007ApJ...655..345J} argue that this model is incompatible with \emph{HST/STIS} observations, especially for the \ion{C}{iv} 1550~\AA{} doublet.
The existence of a hot wind in TW Hya, therefore, remains an open issue.

In this paper we analyse the shifts and widths of hot-ion spectral lines, particularly \ion{C}{iii} 977~\AA{} and the \ion{O}{vi} doublet at 1032~\AA{} and 1038~\AA{}, observed with \emph{FUSE}. This complements earlier studies of all pre-main sequence stars observed with \emph{IUE} \citep{2000ApJS..129..399V}, \emph{GHRS} \citep{2002ApJ...566.1100A,2002ApJ...567.1013A} and of the $H_2$ emission in CTTS \citep{2006ApJS..165..256H}.
In Sect.~\ref{stellarproperties} we briefly summarise the properties of the sample stars, and in Sect.~\ref{observations} we describe the observations and data that we use in our present analysis.  In Sect.~\ref{results} the line profile analysis is presented, several possible formation regions for the observed properties are discussed in Sect.~\ref{discussion}, and in Sect.~\ref{summary} we summarise our results.

\section{Stellar properties}
\label{stellarproperties}

In this section we present the most important properties of our sample stars in Table~\ref{tab_stars} and provide a short summary of their main characteristics.

\subsection{RU~Lupi}
RU~Lupi is a strong accretor, heavily-veiled and strongly variable at different wavelengths \citep{2002A&A...391..595S}. Its period is uncertain. A wind can be studied using cool, metallic lines with a low first-ionisation potential, and in H$_2$ as demonstrated by \citet{2005AJ....129.2777H}.
The line profiles of hot ion lines in the UV exhibit complex structures. \citet{2001MNRAS.323..177T} show by spectro-astrometry that the H$\alpha$ emission is significantly extended and compatible with a bipolar outflow. In X-rays, RU~Lupi is observed to be a typical member of the CTTS class \citep{RULup}; specifically, it shows a low f/i-ratio in the O~VII triplet as well as a soft X-ray excess.

\subsection{T~Tau}
T~Tau is the defining member of the class of TTS. It is a hierarchical triple system, where the northern component \object{T Tau N} is optically visible, while
the southern component \object{T Tau S}, contributing mainly in the infrared, is itself a binary \citep{2000ApJ...531L.147K}. The T~Tau system shows a very
significant soft X-ray excess without the high densities seen in TW~Hya \citep{ttau}. Nevertheless, accretion and winds are detectable using data for other wavelengths \citep{1994A&A...287L..25V,1994ApJ...425L..45V,2005ApJ...625L.131D}, and \citet{2006ApJS..165..256H}, for example, attribute observed fluorescent H$_2$ emission to outflows. 
The first analysis of this FUSE data, completed  by \citet{2002AJ....124.1077W}, provided the first discovery of extra-solar, H2 Werner-band emission.

\subsection{DF~Tau}
DF~Tau is a binary system consisting of two M stars, resolved for the first time by \citet{1990ApJ...357..224C}. The orbital parameters are still uncertain after several revisions \citep{2006AJ....132.2618S} and the distance of the system is still a matter of debate. The most reasonable estimate is provided by \citet{2006A&A...460..499B}, who showed DF~Tau to be a member of the Taurus-Aurigae moving group ($\approx140$~pc), in spite of its low \emph{HIPPARCOS} parallax of $39\pm7$~pc.
\citet{2001AstL...27..313L} present \emph{HST/STIS} and \emph{IUE} spectra showing accretion and wind signatures and attribute DF~Tau's erratic photometric variability to  unsteady accretion. DF~Tau is one of few Doppler-mapped CTTS \citep{1998MNRAS.295..781U}. \citet{2006ApJS..165..256H} demonstrated that fluorescent H$_2$ emission found in \emph{FUSE} observations is consistent with models of a warm, disk surface. X-ray observations for DF~Tau are discussed in Sect.~\ref{obsxray}.

\subsection{V4046~Sgr}
\citet{2000AAS...197.4708S} present the \emph{FUSE} spectrum of V4046~Sgr.
V4046~Sgr is a close binary system with separation of $\approx9R_{\sun}$ \citep{2000IAUS..200P..28Q}. Both of the components are almost identical CTTS, which are accreting matter from a circumbinary disk. \emph{Chandra} X-ray observations show line ratios indicative of high density plasma, which cannot be explained by coronal activity, but must be attributed to hot accretion shocks \citep{v4046}. 

\subsection{TWA~5}
\citet{2000AAS...197.4708S} present the \emph{FUSE} spectrum of TWA~5.
TWA~5 is a hierarchical multiple system, the main component being a M1.5 dwarf. On the one hand, TWA~5 has no significant infrared excess indicating the absence of a disk \citep{2004ApJ...600..435M,2004AJ....127.2246W,2004ApJS..154..439U}; also, X-ray observations by \citet{twa5} suggest that the emission originates in a low-density region, not in an accretion shock. On the other hand, TWA~5 has an H$\alpha$ equivalent width of 13.5~\AA{} and signatures of outflows \citep{2003ApJ...593L.109M}, therefore we regard it as a transition object.

\subsection{GM~Aur}
GM~Aur is one of the most popular systems to observe stellar disk constitution and chemistry \citep[e.g.][]{2004ApJ...614L.133B,2005A&A...442..703H,2007ApJ...655L.105S}. 
Numerous observations in the radio use the disk of GM~Aur to analyse the distribution and evolution of dust grains in the disk \citep[e.g.][]{2006A&A...446..211R,2006A&A...456..535S}. Based on an optically-thin inner disk surrounded by an optically-thick outer region, \citet{2007MNRAS.378..369N} classify GM~Aur as a transition object.
X-ray observations for GM~Aur are discussed in Sect.~\ref{obsxray}.

\subsection{TW~Hya}
Because of its proximity and the absence of a surrounding dark molecular cloud, TW~Hya is a key system for the study of CTTS. Although relatively old by comparison ($\approx10$~Myr), it is still actively accreting from a surrounding disk. 
\citet{2006ApJ...637L.133E} interferometrically resolve its disk and \citet{2004ApJ...607..369H} constrain the inner disk temperature to about 2500~K using H$_2$ fluorescence in \emph{HST/STIS} and \emph{FUSE} data.
Observations with \emph{STIS} and \emph{FUSE} show, in addition to H$_2$, atomic emission lines with notably asymmetric shapes. 
The analytical models expect a cold wind to be driven by magnetic fields from the disk \citep{1994ApJ...429..781S}, and this can be traced observationally from single or double ionised ions \citep{2004AstL...30..413L}. TW~Hya has been extensively observed in X-rays, which revealed a comparatively cool and dense emission region interpreted as signature of an accretion hot spot \citep{2002ApJ...567..434K,twhya}. Our own theoretical modelling supports this view \citep{acc_model}.

\begin{table*}
\caption{\label{tab_stars}Stellar sample properties, N$_{\mathrm{H}}$ lists only values derived from X-ray observations}
\begin{center}
\begin{tabular}{lccccccc}
\hline\hline
Star & Spectral type & d &$v_{rad}$& $v\sin i$ & $i$ & N$_{\mathrm{H}}$ & $A_V$\\
 & & [pc] & [km~s$^{-1}$] & [km~s$^{-1}$]& [$^{\circ}$] & [$10^{21}$~cm$^{-2}$] & [mag]\\
\hline
    RU Lup& G5V & 140 & $-1.9\pm0.02$ (17) & $9.1\pm0.9$ (17) & $24$ (19) & $1.8$ (15) & $0.07$ (9) \\
      T Tau& G5V & 140 & $18$ (9) & $20\pm5$ (9) & $15$ (9) & $5$ (5) & $0.3$ (9) \\
     DF Tau& M0-2& 140 & $21.6\pm9.2$ (4) & $15.8\pm1.0$ (7) & $>60$ (21) & 3-8 & $0.5$ (9)\\
  V4046 Sgr& K5  & 83 & $-6.94$ (12) & $14.9\pm0.9$ (18) & $35$ (18)  & $<2$ (6) & $0.0$ (18)\\
      TWA 5& M3V & 57 & $14$ (20) & uncertain (20) & n.a. & $0.3$ (2) & n.a. \\
     GM Aur& K5V & 140 & $24$ (10) & $13$ (3) & $56\pm2$ (16) & 0.2-0.4 & $0.31$ (11) \\
     TW Hya& K8V & 57 & $12.5\pm0.5$ (1) & $5\pm2$ (1) & $7\pm1$ (13)  & $0.35$ (14) & $0.0$ (8) \\
\end{tabular}
\end{center} 
References
(1) \citealt{2002ApJ...571..378A};
(2) \citealt{twa5};
(3) \citealt{1990ApJ...363..654B};
(4) \citealt{1986A&A...165..110B};
(5) \citealt{ttau};
(6) new fits to the data from \citealt{v4046};
(7) \citealt{1989AJ.....97..873H};
(8) \citealt{2004ApJ...607..369H};
(9) \citealt{2006ApJS..165..256H};
(10) \citealt{1949ApJ...110..424J};
(11) \citealt{1998AJ....116.2965M}
(12) \citealt{2000IAUS..200P..28Q};
(13) \citealt{2004ApJ...616L..11Q};
(14) \citealt{Rob0507};
(15) \citealt{RULup};
(16) \citealt{2000ApJ...545.1034S};
(17) \citealt{2002A&A...391..595S};
(18) \citealt{2004A&A...421.1159S};
(19) \citealt{2007A&A...461..253S};
(20) \citealt{2003AJ....125..825T};
(21) \citealt{1998MNRAS.295..781U}
\end{table*}

\section{Observations}
\label{observations}
\subsection{\emph{FUSE} data}
\label{fusedata}
In order to perform a systematic study of the FUV properties of CTTS, we retrieved all available \emph{FUSE} observations of TTS from the archive.  For a comparison with ``normal" stars, we
selected two main sequence stars, the rapidly-rotating K~star AB~Dor and the solar analog $\alpha$~Cen~A. Table~\ref{tab_obs} lists the observation logs; all targets except $\alpha$~Cen~A are observed using the LWRS aperture with $30''\times30''$ field of view.  We reduced the data with CalFUSE~v3.2.0 \citep{2007PASP..119..527D}, performing extractions of the complete data set and night-only time intervals to check for airglow contamination. With the exception of GM~Aur, the spectral regions of interest are not affected by airglow. 
To coadd the individual exposures, we used the CalFUSE \texttt{get\_shift} task and applied two different methods: 
both methods involved cross-correlation, the first using a stellar emission line (\ion{O}{vi} 1032~\AA{} on the 1aLiF detector and \ion{C}{iii} 977~\AA{} on the 2bSiC detector), and the second method using a strong airglow line (H~Ly$\beta$ and H~Ly$\gamma$). Airglow lines completely cover the aperture and cannot therefore be used to judge centering of a target. We studied all stellar spectra with OVI and CIII sufficiently strong for cross-correlation, and found no significant differences between the line profiles derived using both methods. Using airglow lines does, however, at times enable H$_2$ emission to be detected in the 1bLiF channel. This is in contrast to the less accurate cross-correlation achieved using low S/N stellar lines. 
%We cross-correlated using first a stellar emission line () and second a strong airglow line (). The airglow lines are aperture filling and therefore provide no information about any miscentring of the target star in the aperture, so generally airglow lines are not recommended for wavelength calibration purposes. We checked all stars with \ion{O}{vi} and \ion{C}{iii} lines strong enough for cross-correlation and found no significant differences in the line profiles between the two methods, on the other hand, using the airglow lines allows to detect H$_2$ emission in the 1bLiF channel for some objects, whereas the lines are smeared out by the less accurate cross-correlation using low S/N stellar lines. 

It seems that in our observations the targets did not drift through the apertures. Therefore we shift all airglow H~Ly$\beta$ lines to match the laboratory wavelength. As the targets may be positioned differently in aperture we perform the following check: \citet{2002ApJ...572..310H} discovered molecular hydrogen at rest compared to the star, so we identify H$_2$ lines in the 1bLiF channel in TW~Hya (the shift between channels a and b on the same chip is smaller than 8~km~s$^{-1}$ \citep{2007PASP..119..527D}); the shift of our H$_2$ lines should now match the radial velocity of TW~Hya. This fixes our wavelength scale and from here on we use TW~Hya as our reference spectrum. To avoid the need for a spectral identification, we search the 1bLiF channel of the other observations for strong features, which correlate with TW~Hya. Only lines with the same strengths in full and night-only extractions are considered here. We also check that no airglow feature is expected at that position according to \citet{2001JGR...106.8119F}. The line widths we find are as small as expected \citep{2006ApJS..165..256H}, lending further credibility to the coaddition of single exposures using an airglow line. In addition, we cross-correlate larger regions of the spectra without strong airglow features and find that the required shifts are in agreement with the analysis of single lines. For RU~Lup, T~Tau and DF~Tau we correct for the velocity of the H$_2$ outflow relative to the star from \citet{2006ApJS..165..256H}. Taking into account the stellar radial velocities, the lines shifts confirm the shifts adopted above from the H~Ly$\beta$ lines. For V4046~Sgr, we do not have a \emph{HST/STIS} measurement of H$_2$ lines, and for TWA~5 and GM~Aur no suitable features in the 1bLiF channel were found. However, given the success of the wavelength calibration for other stars, we are quite confident that it is reliable also in these cases. We estimate the total wavelength error in the 1a/bLiF channels to be 20~km~s$^{-1}$.
We then correlate the 2bSiC channel with the 1aLiF channel using the stellar \ion{O}{vi} 1032~\AA{} line. Unfortunately the signal-to-noise ratio (hereafter SNR) in 2bSiC is much lower than in 1aLiF and for T~Tau, DF~Tau and GM~Aur we have to use the H~Ly$\beta$ airglow line. Because the \emph{FUSE} channels are not perfectly aligned, the position of the target in the various channels might be different; the wavelength calibration below 1000~\AA{} for these three stars is therefore uncertain.

For data of low SNR CalFUSE sometimes overcorrects the background, leading to  negative flux intensity values; we therefore always separately fit the background in line-free regions, close to the line of interest.

\begin{table}
\caption{\label{tab_obs}\emph{FUSE} observations}
\begin{center}
\begin{tabular}{lcccc}
\hline \hline
Target 	& Date 		& Exposure time	& Data ID	\\
	&		& [ksec]	& 		\\
\hline
\object{RU Lup} & 2001-08-28 & 	24   & A1090202\\
\object{T Tau}  & 2001-01-15 & 	21   & P1630101\\
\object{DF Tau} & 2000-09-13 & 	26   & A1090101\\
\object{V4046 Sgr}	& 2000-05-18 & 	16   & P1920202\\
\object{TWA 5}	& 2000-05-15 & 	15   & P1920101\\
GM Aur 	  	& 2004-03-03 & 	32   & D0810101\\
\object{GM Aur} & 2004-10-23 & 	34   & D0810102\\
TW Hya 	  	& 2000-06-03 & 	2    & P1860101\\
\object{TW Hya} & 2003-02-20 & 	15   & C0670101\\
TW Hya	    	& 2003-02-21 & 	15   & C0670102\\
%\object{EP Cha}	& 2006-03-21 & 	29   & G0480103\\
%EP Cha		& 2006-03-19 & 	16   & G0480102\\
%EP Cha		& 2006-03-16 & 	18   & G0480101\\
AB Dor	   	& 1999-10-20 & 	22   & X0250201\\
\object{AB Dor} & 1999-12-14 & 	24   & X0250203\\
AB Dor	   	& 2003-12-26 & 	102  & D1260101\\
\object{$\alpha$ Cen A}$^1$ 	& 2001-06-25 & 	15   & P1042601\\
$\alpha$ Cen A$^1$ 	& 2006-05-05 & 	13   & G0810102\\
\end{tabular}
\end{center} 
Notes: (1) MDRS aperture
\end{table}

\subsection{\emph{HST} data}
\label{texthst}
Four of our sample stars (RU~Lup, T~Tau, DF~Tau and TW~Hya) have been observed with gratings onboard of the \emph{HST}. RU~Lup, T~Tau and DF~Tau were observed with the \emph{GHRS}; these observations are presented by \citet{2002ApJ...566.1100A} and \citet{2001AstL...27..313L} and we adopt the linewidths and shifts measured by these authors. The \emph{STIS} observations of RU~Lup were carefully analysed by \citet{2005AJ....129.2777H} and those of TW~Hya by \citet{2002ApJ...572..310H} and \citet{2007ApJ...655..345J}. We use the \emph{STIS} spectra of T~Tau and DF~Tau from CoolCAT \citep{2005ESASP.560..419A}, which is a database of reduced and calibrated UV spectra taken with \emph{STIS} for cool stars. These spectra all belong to the \emph{HST} program 8157 and were taken in the year 2000. The observations of TW~Hya are not contained in CoolCAT, because they were taken in a non-standard setup. For a consistency check, we take T~Tau, and find that the spectra presented there agree with the previously published results in \citet{2005AJ....129.2777H}.

\subsection{X-ray data}
\label{obsxray}
Most of our sample stars have been observed at X-ray wavelengths with {\it Chandra}, \emph{XMM-Newton} or both, and we obtained these data from the literature.
For DF~Tau and GM~Aur, %and EP~Cha 
there are no X-ray spectra available in the literature, however, both DF~Tau and GM~Aur have been observed with \emph{ROSAT/PSPC} pointings.
We therefore retrieved the \emph{ROSAT} X-ray data from the archive in order to enlarge our sample for comparisons of optical and X-ray absorption.
In Table~\ref{tab_rosat}, we provide information on the data used. The \emph{ROSAT} data were reduced using EXSAS version 03Oct. Source photons were extracted
from a circle with a radius of 2.5 arcmin to avoid contamination from nearby sources. The background was taken from a region close to the target without detected sources with a radius three times larger. 
We measured fluxes of $0.013\pm0.002$~cts~s$^{-1}$ for DF~Tau and $0.021\pm0.003$~cts~s$^{-1}$ for GM~Aur.
Using MIDAS/EXSAS, we fitted Raymond-Smith and Mewe-Kaastra models assuming cold absorption. 
The ROSAT data are of low SNR. We therefore tried to use different data binnings and compared our final results.
We provide a range of possible absorption column-densities in Table~\ref{tab_stars}, because, as often found, observationally a large amount of cool plasma and small interstellar absorption lead to very similar signatures.
%In the fit process there is -- as always -- an ambiguity between the amount of cool plasma and interstellar absorption and 
%we therefore give a range of possible values for the absorption column density in Table.
To test the reliability of our procedure,
we fit the \emph{ROSAT} data for AB~Aur, which is located in the same field as GM~Aur. We measure an absorption column-denstiy  N$_{\mathrm{H}}$ of $2-3\times10^{20}$~cm$^{-2}$, which is approximately 
half of the measurement derived using \emph{XMM-Newton} data presented by \citet{ABAur}.  We therefore conclude that, taking into account systematic errors, the \emph{ROSAT}-derived absorption columns are consistent with those derived using \emph{XMM-Newton} data.

\begin{table}
\caption{\label{tab_rosat}\emph{ROSAT} observations}
\begin{center}
\begin{tabular}{lcccc}
\hline \hline
Target 	& Date 		& Exposure time	& Obs ID	\\
	&		& [ksec]	& 		\\
\hline
DF Tau	& 1993-08-93	& 10		& WG201533P.N1\\
GM Aur	& 1992-09-22	& 1.3		& WG201278P.N1\\
GM Aur	& 1993-03-04	& 4		& WG201278P-1.N1\\
\end{tabular}
\end{center} 
\end{table}

\section{Results}
\label{results}
\subsection{FUV lines}

\begin{figure}
\resizebox{\hsize}{!}{\includegraphics{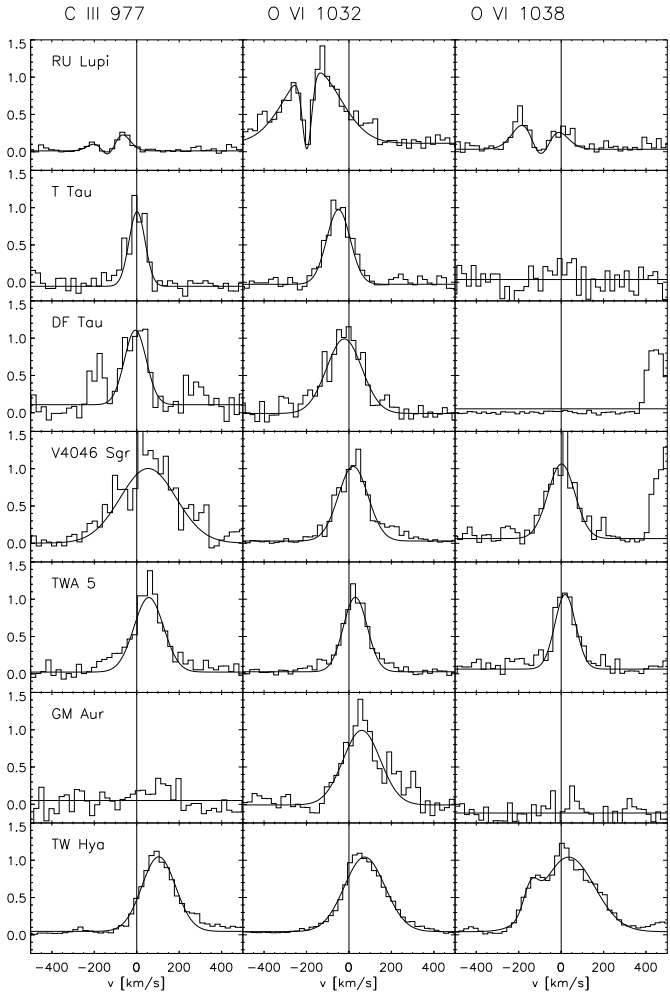}}
\caption{\label{lines}Hot ion lines observed with \emph{FUSE} in CTTS and best fit Gaussian profiles ordered by the shift in the \ion{O}{vi} 1032~\AA{} line. All line profiles are normalised to their the Gaussian profile peak value and rebinned to instrumental resolution.}
\end{figure}

In Fig.~\ref{lines}, we present the \ion{C}{iii} line profiles  at 977~\AA{} and both components of the \ion{O}{vi} doublet 
at 1032~\AA{} and 1038~\AA{} for our targeted stars. Each spectral line is fitted with a Gaussian profile and shown
normalised with respect to the fitted peak value. The best-fit parameters for line flux, shift of the Gaussian-line centre, and its full width at half maximum (FWHM), ordered by the observed shift in the \ion{O}{vi} 1032~\AA{} line, are given in Table~\ref{tab_gaussfits}, where the shift is corrected for the radial velocity of the star. The statistical errors  on the shift and the FWHM given in the table are typically comparable to our systematic calibration uncertainties. The systematic errors on the fluxes are of the order 10-20\%.
The \ion{O}{vi} doublet line at 1032~\AA{} is detected in all sample stars  and the \ion{C}{iii} line in all except GM~Aur. In all cases, the line shifts measured in the \ion{C}{iii} and \ion{O}{vi} are consistent, arguing for related emission regions; only for T~Tau is there a statistically significant difference, probably due to uncertainties in the wavelength calibration of 2SiC detector data, as discussed in Section~\ref{fusedata}.

In the case of RU~Lupi, an absorption component is clearly visible in all three lines, which is also fitted with a Gaussian absorption line. We note that this absorption is not self-absorption. \citet{2005AJ....129.2777H} identify the absorption in the \ion{O}{vi} 1032~\AA{} as H$_2$ 6-0~P(3) shifted by $-23\pm5$~km~s$^{-1}$ relative to the H$_2$ rest wavelength, and the absorption in \ion{O}{vi} 1038~\AA{} as \ion{C}{ii}. Using the package H2oolts \citep{2003PASP..115..651M}, we identify the absorption component in \ion{C}{iii} to be a superposition of the equally strong H$_2$ 2-0~P(5) and 11-0~R(3) shifted by $\approx-10\pm15$~km~s$^{-1}$ again relative to the H$_2$ rest wavelength.
Within the \ion{O}{vi} 1038~\AA{} line of TW~Hya, we fit the \ion{C}{ii} blend on the blue side with an additional Gaussian component.
Unfortunately the data quality does not enable further blends and features to be identified. In the \emph{FUSE} bandpass, numerous  fluorescent Lyman and Werner band H$_2$ lines are present, which are usually relatively sharp \citep{2006ApJS..165..256H}. The lines in Fig.~\ref{lines} are well-described by one-peaked distributions, so the contribution of any individual blending line has to be small compared to the broad ion lines. 

\setlength{\tabcolsep}{4.3pt}
\begin{table*}
\caption{\label{tab_gaussfits} \emph{FUSE} best fit parameters, fluxes in 10$^{-15}$~erg~cm$^{-2}$~s$^{-1}$, width and shift in km~s$^{-1}$, errors are statistical only}
\begin{center}
\begin{tabular}{l|ccc|ccc|ccc|c}
\hline\hline
 & \multicolumn{3}{c|}{\ion{C}{iii} 977 \AA{}} & \multicolumn{3}{c|}{\ion{O}{vi} 1032 \AA{}} & \multicolumn{3}{c|}{\ion{O}{vi} 1038 \AA{}} & \ion{C}{iii} 1175~\AA{}\\
Star & Flux & Shift & FWHM & Flux  & Shift & FWHM & Flux & Shift  & FWHM & Flux\\ 
% & & [km~s$^{-1}$] & [km~s$^{-1}$] & & km~s$^{-1}$ & km~s$^{-1}$ &  & km~s$^{-1}$& km~s$^{-1}$ \\
\hline
  RU Lup& $   70\pm   60$& $ -120\pm   10$& $  120\pm   20$& $   24\pm    6$& $ -170\pm   20$& $  300\pm   60$& $   15\pm    8$& $ -100\pm   20$& $  170\pm   30$ & $20\pm2$\\
-- Absorption& $  -60\pm   40$& $ -130\pm   10$& $  100\pm   20$& $   -4\pm    0$& $ -200\pm    6$& $   50\pm   20$& $  -10\pm    2$& $ -100\pm   10$& $  120\pm   30$ & ...\\
    T Tau& $   10\pm    4$& $    1\pm   13$& $   90\pm   30$& $    8\pm    2$& $  -50\pm   10$& $  130\pm   30$&  n.a. &    n.a. &   n.a.  & $18\pm2$ \\
   DF Tau& $   17\pm    6$& $   -6\pm   12$& $  120\pm   30$& $   11\pm    3$& $  -22\pm   16$& $  200\pm   40$&    n.a. &   n.a. &   n.a.  & $44\pm2$ \\
V4046 Sgr& $   60\pm   20$& $   50\pm   20$& $  310\pm   70$& $   37\pm    6$& $   19\pm    8$& $  160\pm   20$& $   16\pm    4$& $    1\pm   12$& $  150\pm   30$ & $50\pm3$\\
    TWA 5& $   30\pm   10$& $   60\pm   20$& $  160\pm   40$& $   30\pm    5$& $   30\pm    7$& $  130\pm   20$& $   14\pm    3$& $   19\pm    9$& $  110\pm   20$ & $25\pm2$\\
   GM Aur&   n.a. &   n.a. &   n.a. & $    5\pm    1$& $   60\pm   20$& $  210\pm   60$&    n.a. &   n.a. &   n.a. & $24\pm2$ \\
   TW Hya& $  200\pm   25$& $  100\pm    6$& $  190\pm   20$& $  220\pm   20$& $   74\pm    5$& $  220\pm   15$& $  100\pm   15$& $   35\pm   15$& $  290\pm   40$ & $310\pm4$\\
-- Blend & ... & ... & ... & ... & ... & ...& $    9\pm    9$& $ -140\pm   14$& $   80\pm   40$ & ...\\ 

\end{tabular}
\end{center}
\end{table*}
\setlength{\tabcolsep}{6pt}	% 6pt is standard

Atomic physics predicts that the 1032 \AA{} component of the \ion{O}{vi} doublet should be twice as strong as the 1038 \AA{} component.
We can confirm this prediction for the data of V4046~Sgr and TWA~5. 
In Table~\ref{tab_gaussfits}, significant deviations from the expected line ratio can be seen in RU~Lup and TW~Hya, where additional components complicate the fit.  However, in the cases of
T~Tau, DF~Tau and GM~Aur the \ion{O}{vi} 1038~\AA{} line is almost completely absent, because of absorption by larger molecular hydrogen columns (see Table~\ref{tab_stars}) in the lines H$_2$ 5-0~R(1) at 1037.1~\AA{} and 5-0~P(1) at 1038.2~\AA{}. RU~Lup and T~Tau, both CTTS with blueshifted emission, are found to have the largest reddening and absorption column density.

\begin{figure}
\resizebox{\hsize}{!}{\includegraphics{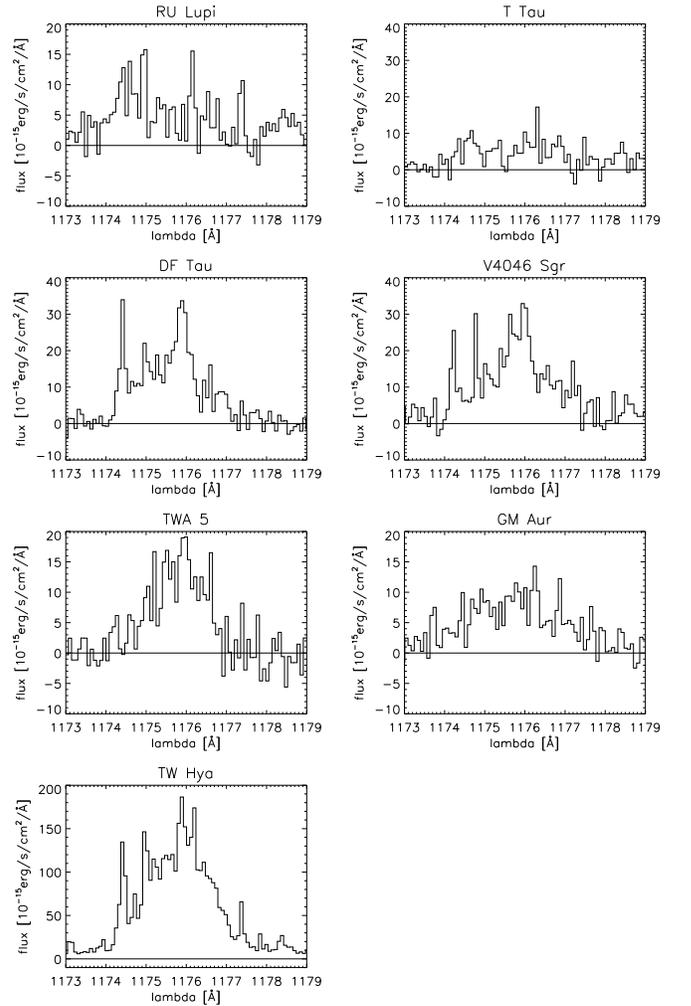}}
\caption{\label{c1175}The \ion{C}{iii} multiplet around 1175~\AA{} rebinned to instrumental resolution.}
\end{figure}

In Fig.~\ref{c1175}, we present the FUSE observations of the \ion{C}{iii} multiplet at wavelength 1175~\AA{}. Unfortunately the SNR of the data is too low to be able to resolve the H2 lines, and to fit the individual line components. This analysis would have enabled density diagnostics to be measured. For RU Lup and T Tau in fact, the multiplet is barely detectable. Therefore a detailed Gaussian function fit yields no useful information, and in Table~\ref{tab_gaussfits} we present only the total flux and its statistical error.  The dereddened ratio of the fluxes in the \ion{C}{iii} 977~\AA{} line and the 1175~\AA\  multiplet is, in principle, density-sensitive in the range $8<\log n<11$ according to the CHIANTI database \citep{CHIANTII,CHIANTIVII}.  However,
apart from RU~Lup the observed values are far below the predicted ratios. \citet{2002ApJ...581..626R} have analysed this line ratio in \emph{FUSE} observations of late-type dwarf stars, and \citet{2005ApJ...622..629D} for luminous cool stars. Both authors conclude that the \ion{C}{iii} 977~\AA\ line is optically-thick in the transition region and that in addition there is  strong interstellar line absorption.

\begin{figure}
\resizebox{\hsize}{!}{\includegraphics{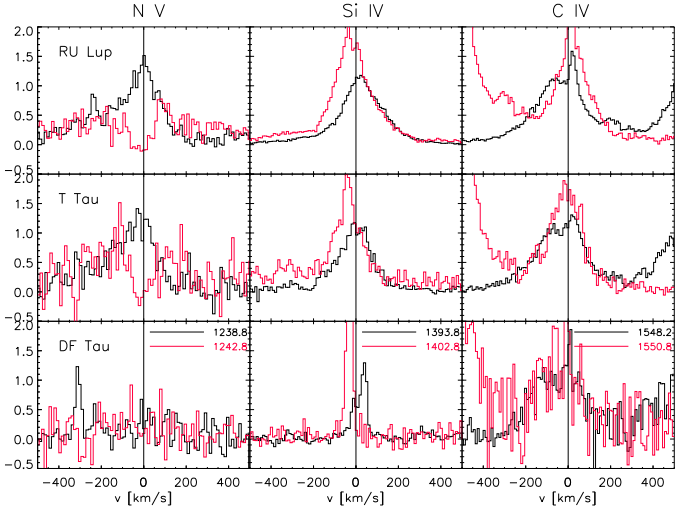}}
\caption{\label{stis}Hot ion lines in the \emph{STIS} spectra. The relative flux of the red doublet member is multiplied with a factor of 2 (red/grey lines).}
\end{figure}

The \emph{STIS} data from CoolCAT is shown close to the hot ionic doublets \ion{N}{v} 1240~\AA{}, \ion{Si}{iv} 1400~\AA{} and \ion{C}{v}~1550~\AA{} in Fig.~\ref{stis}.
In each case, the expected intensity ratio is two to one. To facilitate the comparison of the line profiles we show the flux of the red doublet member increased by a factor of two. In the absence of blending and interstellar absorption, both lines should show a perfect match. The red member of the \ion{N}{v} doublet at 1242.8~\AA{} is totally absorbed by an interstellar \ion{N}{i} line, so we do not attempt to fit the line profile. For the other lines, the results of a fit with a single gaussian component are shown in Table~\ref{HST}. In DF~Tau there is no overlap between the two lines shown in the middle panel of Fig.~\ref{stis}. Therefore they cannot be doublet members, but have to be other lines, likely H$_2$. These lines also distort the red member of this doublet in the other two stars, leading to a shifted centre in the gaussian fit. The large errors result from generally non-Gaussian line shapes. The Table also includes the \emph{GHRS} data and the observations of TW~Hya, resembling the \emph{FUSE} data. These profiles are nearly triangular, rising sharply close to the rest wavelength and extending to about 400~km~s$^{-1}$ on the red side \citep[see Fig.~7 in][]{2002ApJ...572..310H}. Of similar shape, although extending only to 200~km~s$^{-1}$ on the red side, is the \ion{C}{iv} 1550~\AA{} doublet in DF~Tau \citep{2001AstL...27..313L} in the \emph{GHRS} observation. This data is fitted with multiple Gaussians by \citet{2002ApJ...566.1100A}. We give here the values for the dominant component only.
\setlength{\tabcolsep}{4.3pt}
\begin{table*}
\caption{\label{HST} \emph{HST} best fit parameters, for references see Sect.~\ref{texthst}.}
\begin{center}
\begin{tabular}{l|l|cc|cc|cc|cc|cc}
\hline\hline
& & \multicolumn{2}{c|}{\ion{N}{v} 1239 \AA} & \multicolumn{2}{c|}{\ion{Si}{iv} 1394 \AA} & \multicolumn{2}{c|}{\ion{Si}{iv} 1403 \AA}& \multicolumn{2}{c|}{\ion{C}{iv} 1548 \AA} & \multicolumn{2}{c}{\ion{C}{iv} 1551 \AA}\\
Star & Instr. & Shift & FWHM  & Shift& FWHM & Shift & FWHM & Shift & FWHM  & Shift& FWHM\\
 & & km~s$^{-1}$ & km~s$^{-1}$ & km~s$^{-1}$ & km~s$^{-1}$ & km~s$^{-1}$ & km~s$^{-1}$ &  km~s$^{-1}$ & km~s$^{-1}$ & km~s$^{-1}$ & km~s$^{-1}$\\
\hline
   RU Lup& GHRS & ... & ... & $    5\pm    5$& $  270\pm   10$& $  -31\pm    6$& $  250\pm   10$& $  -20\pm   20$& $  340\pm   40$& $  -10\pm   10$& $  300\pm   30$ \\
   RU Lup& STIS & $  -15\pm   25$& $  190\pm   70$& $   26\pm   20$& $  200\pm   50$& $  -20\pm   20$& $  180\pm   50$& $  -25\pm   25$& $  230\pm   80$& $   15\pm   18$& $  120\pm   50$ \\
    T Tau& GHRS & ... & ... & $  -15\pm    7$& $  220\pm   20$& $  -50\pm   10$& $  220\pm   30$& $  -25\pm    6$& $  250\pm   10$& $    0\pm    6$& $  150\pm   10$ \\
    T Tau& STIS & $  -45\pm   30$& $  220\pm   80$& $   -1\pm   20$& $  180\pm   50$& $  -40\pm   20$& $  130\pm   50$& $  -25\pm   20$& $  230\pm   70$& $  -10\pm   20$& $  150\pm   50$ \\
   DF Tau& GHRS & ... & ... & ... & 300 & ... & 300 & $   10\pm   10$& $  350\pm   30$& $   10\pm   10$& $  320\pm   20$ \\
   DF Tau& STIS & $   30\pm   40$ & $  187\pm   85$ & ... & ... & ... & ... & $  -40\pm   40$& $  230\pm  100$& $  -50\pm   40$& $  180\pm  120$ \\
   TW Hya& STIS & \multicolumn{2}{c|}{redshifted, see text}& $   42\pm    3$& $  170\pm    6$& $    9\pm    3$& $  130\pm    8$& \multicolumn{4}{c}{redshifted, non-gaussian, see text} \\
\end{tabular}
\end{center}
\end{table*}
\setlength{\tabcolsep}{6pt}
 
Despite the uncertainties in the profiles of individual lines, it is obvious from Tables~\ref{tab_gaussfits} and \ref{HST} and Fig.~\ref{lines} and \ref{stis}, that FUV lines in CTTS cover a range of wavelength shifts from blue-shifted to red-shifted velocities of the order of -200~km~s$^{-1}$ out to 100~km~s$^{-1}$, and could therefore potentially originate in in- or outflows or both. RU~Lup clearly shows blue-shifted emission, the same is likely the case for T~Tau, because the \ion{O}{vi} shift can be more accurately determined than the \ion{C}{iii} shift. DF~Tau and V4046~Sgr are consistent with centred lines and TWA~5, GM~Aur and TW~Hya with red-shifted lines.

\subsection{Excess absorption}
\begin{figure}
\resizebox{\hsize}{!}{\includegraphics{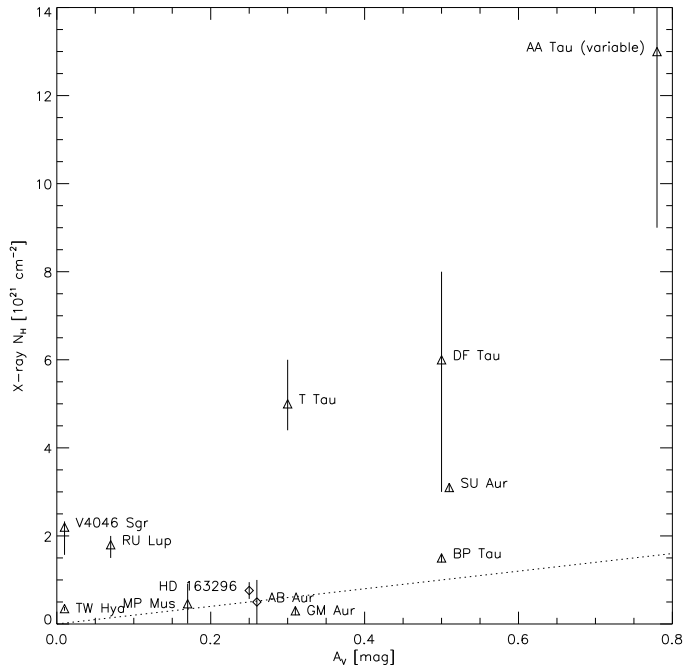}}
\caption{\label{nhav}Optical reddening and absorbing column density from X-rays. Error bars represent 90\% confidence intervals on the $N_H$ fit, except for DF~Tau and GM~Aur where we give the (tighter) ranges discussed in Sect.~\ref{obsxray}. The dotted line indicates the values predicted using $N_H=A_V \cdot2\times 10^{22}\mathrm{cm}^{-2}$.}
\end{figure}

In Fig.~\ref{nhav}, we show the absorbing column densities as fitted to the X-ray observations and optical $A_V$ values. With a standard gas-to-dust ratio, they should be related through the formula
$ N_H=A_V \cdot2\times 10^{22}\mathrm{cm}^{-2}$
\citep[][but see \citet{2003A&A...408..581V} for a compilation of other conversion factors in the literature, all roughly consistent with this value]{1979ARA&A..17...73S}.
The determination of $A_V$ values is difficult for CTTS, where not only the spectral type, but also the veiling needs to be known to compute the intrinsic colours. This renders $A_V$ values in the literature notoriously uncertain, especially if they are based only on photometric and not on spectroscopic information. We have tried to compile the best available estimates in Table~\ref{tab_stars}. For \object{AA~Tau} and possibly other CTTS, the circumstellar absorption and the veiling are time-dependent. In Fig.~\ref{nhav}, we show mean values for AA~Tau from \citet{2007A&A...462L..41S} and \citet{1999A&A...349..619B}. The $A_V$ for T~Tau is often given one magnitude larger than the value we use, but \citet{2003AJ....126.3076W} convincingly show that this is an overestimate. In the case of DF~Tau, most values are between 0.21 \citep{1995ApJS..101..117K} and 0.55 \citep{1989AJ.....97.1451S}, with the exception of $1.9\pm0.6$ in \citet{1979ApJS...41..743C}; for RU~Lup, all modern measurements agree on the low $A_V$ value \citep{1996A&A...306..877L,2002A&A...391..595S,2005AJ....129.2777H}.
In addition to the TTS studied in this paper, we included data for the young stars \object{BP~Tau} and \object{SU~Aur} from \citet{Rob0507} and \citet{2001A&A...377..557E}, for BP~Tau, \object{AB~Aur} and \object{HD~163296} from \citet[][see references therein]{ABAur} and for \object{MP~Mus} from \citet{2007A&A...465L...5A} and \citet{2002AJ....124.1670M}.
In the case of AA~Tau, reddening and absorbing column are exceptionally large, because this object is seen almost edge-on. AA~Tau, DF~Tau, T~Tau, RU~Lup and V4046~Sgr all show much larger X-ray absorbing columns than expected from the optical extinction, indicating large amounts of gas with a very small dust content compared to the interstellar medium. This effect is significantly larger than the uncertainties in the $A_V$ and $N_H$ values. In V4046~Sgr the relative uncertainties are higher compared to the other stars, because \emph{Chandra} is not as sensitive as \emph{XMM-Newton} at soft X-ray energies.
In RU~Lup and, possibly, T~Tau the hot ion lines exhibit significant blue shifts. Therefore, it may well be that the hot UV-emitting plasma provides the observed X-ray excess absorption.

All $N_H$ values provided are obtained by fitting a cold absorber model, implying that the elements contributing to the X-ray absorption are not fully ionised. The most important elements are O, H, and He; with increasing temperature and ionisation an even larger column density is necessary to provide the observed absorption.

\subsection{Ratio of \ion{O}{vii} to \ion{O}{vi} luminosity}

\begin{figure}
\resizebox{\hsize}{!}{\includegraphics{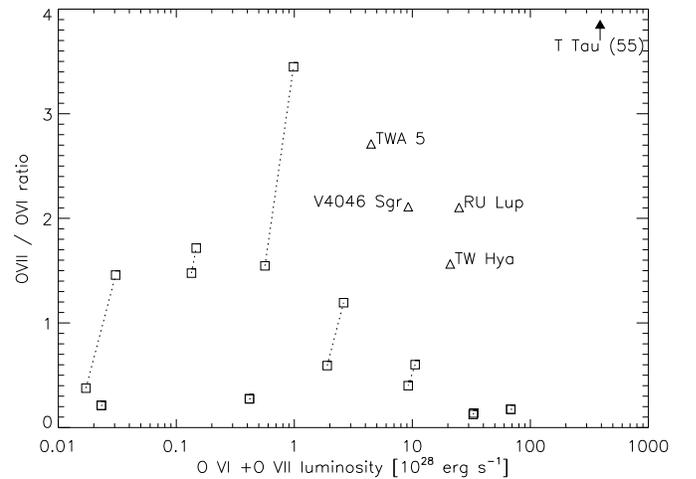}}
\caption{\label{o62o7}Ratio of dereddened fluxes in the \ion{O}{vii} triplet at 22~\AA{} to the \ion{O}{vi} 1032~\AA{} line vs. total luminosity in those lines for CTTS (triangles) and other stars (squares). See text for data sources.}
\end{figure}

X-ray data are available for all CTTS sample stars, 
 most of which were observed using gratings. The resolution
of these X-ray lines is insufficient to determine velocities, and we therefore analyse only the measured line fluxes.
In Fig.~\ref{o62o7}, we compare the fluxes of the \ion{O}{vi} and \ion{O}{vii} lines, where the X-ray data is taken from the same references as N$_{\mathrm{H}}$ in Table~\ref{tab_stars}. The observed fluxes are dereddened \citep{1989ApJ...345..245C} in the UV with a standard dust grain distribution assuming $R_V=3.1$, and in the X-rays following the cold absorption model by \citet{1992ApJ...400..699B}.
For comparison, some stars with ''normal`` coronal X-ray emission have been added \citep{2004A&A...427..667N}, whose \ion{O}{vi} emission is known \citep{2002ApJ...581..626R,2005ApJ...622..629D}. Where \citet{2004A&A...427..667N} provide measurements from both \emph{XMM-Newton} and \emph{Chandra}, we show two connected symbols to give an idea of the intrinsic variability of the sources. This by far dominates the observational uncertainties, which are approximately 10\%.
As is clear from Fig.~\ref{o62o7},
the CTTS separate well out in the upper right corner of the plot, indicating an \ion{O}{vii} excess compared to dwarf and giant cool stars. 
\citet{RULup} and \citet{manuelnh} show a very similar analysis using the \ion{O}{viii}/\ion{O}{vii} ratio again with an \ion{O}{vii} excess. 
These studies clearly show that the observed excess radiation of CTTS is confined to a relatively narrow temperature range about the formation of the He-like \ion{O}{vii} triplet at 1-2~MK. We note that
the CTTS T~Tau has an unusually strong \ion{O}{vii} excess, with a ratio close to 55.
Estimating the H$_2$ column density to be roughly a quarter of the total neutral hydrogen absorbing column \citep{1987ApJ...322..694B}, no more then about a third of the \ion{O}{vi} flux in T~Tau could be absorbed and the true blueshift cannot exceed 100~km~s$^{-1}$. Otherwise emission blueward of the absorbing H$_2$~P(3)(6-0) line at 1031.19~\AA{} would be observed in Fig.~\ref{lines}. Absorption in the \ion{O}{vi} lines cannot therefore explain the extreme \ion{O}{vii} to \ion{O}{vi} ratio. A possible origin for the large emission measure at 1-2~MK might be a small-density accretion shock with a large filling factor.

\section{Interpretation}
\label{discussion}
In the following we attempt to constrain the emission regions and set limits and constraints 
on possible models for the origin of the observed line shapes.

\subsection{Blue-shifted emission}
The detected emission of hot-ion lines in the \emph{FUSE} data appears blue-shifted in RU~Lupi and, probably, in T~Tau. Both systems show small values of $v\sin i$ compared to zero-age main-sequence stars and their inclination is close to face-on (Table~\ref{tab_stars}). However, TW~Hya is a counter-example with small inclination and red-shifted emission.

\subsubsection{Shocks in outflows}

We assume that an outflow exists, produced by either a disk or stellar wind.
Blue-shifted emission could then be due to either internal shock fronts or to termination shocks of outflows running e.g. into an
ISM cloud. For the case of RU~Lupi, the measured centroid velocity is found to be $v_{\mathrm{obs}}\approx200~\mathrm{km~s}^{-1}$ away from the star, and the peak formation temperature of \ion{O}{vi} is $300\,000$~K \citep{mazzottaetal}.
If we assume that the \ion{O}{vi} gas is heated by a strong shock wave, following \citet{raizerzeldovich} we can infer that the shock front properties 
from the Rankine-Hugoniot jump conditions. The emission of a post-shock cooling zone would be observed with velocity $v_{\mathrm{obs}}$, which sets a lower boundary on the shock velocity $v_{\mathrm{shock}}$ to 
\begin{equation} 4 v_{\mathrm{obs}}=v_{\mathrm{shock}} \ge 800\mathrm{km~s}^{-1} \ . \label{vstrongshock}\end{equation} 
By using line shifts we measure only the line of sight velocity, but RU~Lupi and T~Tau are probably viewed close to pole-on, thus
we expect this value to be close to the true velocity.  Typical jets reach up to 400~km~s$^{-1}$ \citep{1998AJ....115.1554E}, but \citet{1995RMxAC...1....1R} report velocities of up to 1000~km~s$^{-1}$. The highest bulk velocities observed in the jet of the CTTS \object{DG Tau} are about 500~km~s$^{-1}$ \citep{2000A&A...356L..41L}.
Using the strong shock formula the post-shock temperature $T_{post}$ can be obtained by 
\begin{eqnarray} T_{\mathrm{post}} & =& \frac{\gamma-1}{(\gamma+1)^2}\frac{2\mu m_{\mathrm{H}}}{k}v_{\mathrm{shock}}^2 \label{T_schock_general}\\
 & = & \frac{3\mu m_{\mathrm{H}}}{k}v_{\mathrm{obs}}^2 \nonumber \ ,\end{eqnarray}
where $m_{\mathrm{H}}$ is the mass of an hydrogen atom, $\mu$ the dimensionless mean particle mass, $k$ is Boltzmann's constant and $\gamma$ is the adiabatic index. In the last step, we use $\gamma=5/3$ for an ideal gas and Eqn.~\ref{vstrongshock}. The estimated post-shock temperatures of $\approx 2\times10^6$~K for RU~Lup and $\approx 2\times 10^5$~K for T~Tau are large enough to form \ion{O}{vi} and \ion{O}{vii}. This seems plausible, because \citet{2005ApJ...626L..53G} present already spatially-resolved X-ray emission from a CTTS jet in DG~Tau~A.
The shifts of the \ion{O}{vi} lines observed in RU Lup and T Tau are larger than measured for the \ion{C}{iii} lines, agreeing with models of post-shock cooling.
\ion{C}{iii} is formed at lower temperatures, when the post-shock flow has lost energy, cooled and slowed down. 

The observed volume emission measures (in units of $10^{51}$~cm$^{-3}$)
for RU~Lup, T~Tau and DF~Tau in the \ion{O}{vi} 1032~\AA{} and 1038~\AA{} lines are 1.3, 1.4 and 4.7, assuming the temperature of maximum emissivity and dereddening of fluxes by \citet{1989ApJ...345..245C}; 
for \ion{C}{iii} the corresponding numbers are 3.9, 1.5 and 8.0. 
Along the jet of the CTTS \object{DG~Tau} pre-shock densities between $10^3$ and $10^5$~cm$^{-3}$ are observed, with the higher values closer to the star \citep{2000A&A...356L..41L}. 
We now use the fit-formula from \citet{1987ApJ...316..323H} to estimate the cooling distance $d$ and we find $d\approx3\times10^{16}$~cm for RU~Lup with $v=800\mathrm{~km~s}^{-1}$ and $n=10^4$~cm$^{-3}$. The diameter of the jet is then of the order $10^{13}$~cm and the column density of the cooling zone alone $N_H=3\times10^{20}$~cm$^{-2}$, adding to the absorbing column in the base of the jet, the circumstellar material further out and the interstellar column density. Therefore the large column densities observed for the stars with blue-shifted emission might be partly due to jet material. In T~Tau and DF~Tau the observed velocities are lower, which results in $d\approx3\times10^{14}$~cm. This is too small to explain the observed emission measures. In the jet emission, picture this suggests jets inclined to the line of sight, so the velocities and therefore the cooling distances are underestimated. According to the data in Table~\ref{tab_stars}, the inclination of T~Tau is smaller than the inclination of RU~Lup, but it is not clear if jets are always orientated perpendicular to the accretion disk. DF~Tau has the largest inclination, so we expect to see the jet at a high inclination. This might explain the large emission measure despite the low observed shift, which is even consistent with centred emission.

It could thus well be that the additional absorption of the stellar X-ray emission is due to the hot gas from which the FUV lines originate. 

\subsection{Anomalous line widths}

\begin{figure}
\resizebox{\hsize}{!}{\includegraphics{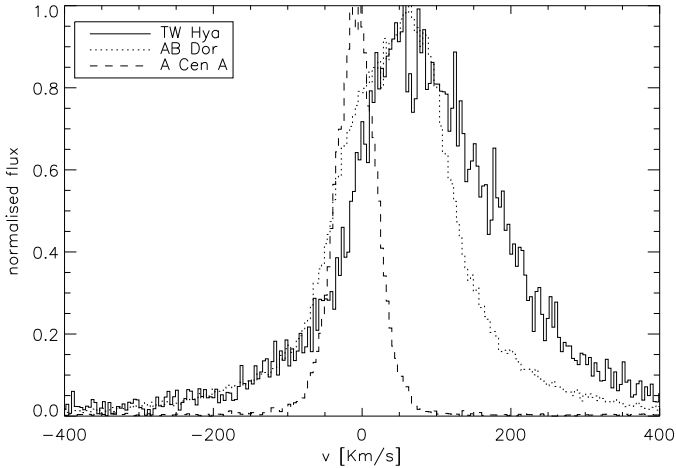}}
\caption{\label{linecomp}Comparison of the \ion{O}{vi} 1031.91~\AA{} line for three stars.}
\end{figure}

A comparison of rotational velocities listed in
Table~\ref{tab_stars} with the line widths shown in Table~\ref{tab_gaussfits} and \ref{HST} shows that the observed lines are superrotationally-broadened despite the measurement uncertainties. This is not uncommon in late-type stars \citep{1997ApJ...478..745W}.
To illustrate this issue further, in Fig.~\ref{linecomp} we compare the \ion{O}{vi} 1032~\AA{} line measured for TW~Hya, the much more rapidly rotating K dwarf AB~Dor with $v\sin i=100\pm5$~km~s$^{-1}$ \citep{1987ApJ...320..850V}, and the slow-rotator
$\alpha$~Cen with $v\sin i=2.7\pm0.7$~km~s$^{-1}$ \citep{1997MNRAS.284..803S}
is in fact comparable to TW~Hya. \citet{2000ApJ...538L..87A} found significantly superrotationally-broadened coronal lines already in AB~Dor and speculate that this might be due to microflare heating. The blue wing of TW~Hya matches the data from AB~Dor quite well, but the asymmetric profile from TW~Hya is significantly different on the red side. At any rate, the  \ion{O}{vi} 1032~\AA{} is clearly far broader than the line from $\alpha$~Cen.  In CTTS, redward line asymmetries are usually attributed to emission in the accretion process. We therefore consider if this interpretation is also valid for TW~Hya. 

\subsection{Magnetospheric infall}
\label{model}
\subsubsection{The basic model}

In the standard accretion scenario for CTTS, the disk is truncated at a few stellar radii and the temperatures on the disk surface are a few thousand~K at most \citep{2004ApJ...607..369H}. Consequently, excluding additional energy deposition by magnetic star-disk interaction, the accreting material reaches the formation temperature of \ion{O}{vi} for the first time shortly before it passes through the accretion shock. At this stage, the material on the opposite side of the star ought to be blocked from view and only the low velocity, low temperature zones of the accretion funnel, close to their origin on the disk, can be seen blue-shifted if they face the observer. Therefore hot ion lines can be observed only red-shifted from the funnel flow.

Using the CHIANTI 5.2 database \citep{CHIANTII,CHIANTIVII} we have numerically modelled the accretion spot as described in detail by \citet{acc_model} for TW~Hya and for V4046~Sgr \citep{Bonn06}. We measure the infall velocities in the accretion funnel to be 525~km~s$^{-1}$, and the pre-shock densities around $10^{12}$~cm$^{-3}$ for TW~Hya and $2\times10^{11}$~cm$^{-3}$ for V4046~Sgr. This matter passes through a shock on the stellar surface and is heated up to temperatures of $\approx 2\times10^6$~K. X-ray emission is produced in the following cooling flow in the post-shock zone. We explicitly calculate the ionisation and recombination rates, to be able to assess the ionisation state at each depth. 
Fitting this model to high-resolution X-ray data, we measure the abundances of the most important elements and determine mass accretion rates and filling factors between 0.1\%-0.4\%. 

Our model explicitly assumes optically thin-radiation loss. Along the direction of infalling flow, most resonance lines are found to be optically-thick up to optical depths above 10. The opacity is measured using a simple opacity calculation, and its precise value depends on both the infall density and elemental abundance.
In the He-like triplets of \ion{Ne}{ix} and \ion{O}{vii}, observed in X-rays, the ratio of resonance line to intercombination plus forbidden line is expected to be close to unity from atomic physics. This is matched by observations, so we can infer that the resonance lines do not suffer much absorption. 
We note that the detailed accretion region geometry is unknown.  In particular, the system geometry is probably not 1-D and
photons may well escape through the boundaries of the system rather than along the direction of flow. 
Unfortunately, this leads to considerable uncertainty in the computed flux. In our simulation however, emission is generated in the post-shock region to the same extent as observed for the OVI triplet, but the post-shock contribution to the CIII line is small.
We caution that this conclusion is based on the assumption of very small and hot spots.
Theoretical simulations of the infall geometry using either dipolar fields \citep{2004ApJ...610..920R} or more realistic geometries scaled from other stars \citep{2006MNRAS.371..999G} suggest inhomogeneous spots with a distribution of infall velocities. 
In accretion zones with smaller infall velocities, the post-shock temperatures and hence the emitted X-ray fluxes are smaller.
Such effects are not included in our models, but could contribute to the observed FUV flux. 
We are unable to reliably constrain fits of inhomogeneous spots due to insufficient diagnostics.

\subsubsection{Predicted UV line shifts}
For TW~Hya and V4046~Sgr, where we modelled the accretion shock in detail, the largest possible post-shock speed is 125~km~s$^{-1}$, a quarter of the free-fall velocity. All emission redward of this value must originate in the pre-shock radiative precursor.

\begin{figure}
\resizebox{\hsize}{!}{\includegraphics{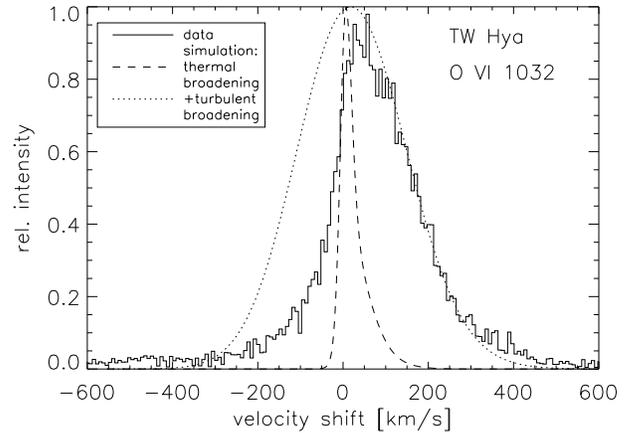}}
\caption{\label{sim_line_profile}\protect{\ion{O}{vi}} line at 1032~\AA{}. solid line: FUSE data, dashed line: simulation with purely thermal broadening, dotted line: simulation with additional turbulent broadening of 150~km~s$^{-1}$}
\end{figure}

During the cooling behind the shock front, the velocity decreases further.
The emission from \ion{C}{iii} and \ion{O}{vi} peaks at very low velocities. 
For pre-shock velocities of 300~km~s$^{-1}$, \ion{O}{vi} is formed immediately behind the accretion shock. Its velocity reaches a maximum value of 75~km~s$^{-1}$, and decreases as the cooling proceeds. Depending on the stellar inclination and the spot latitude, the projected velocity may be even smaller. The resulting line profile has its maximum close to $v=0$~km~s$^{-1}$ with a broad wing
on the red side.  Observationally the \ion{O}{vi} line at 1032~\AA{} is
best suited for analysis, because the other member of the doublet at 1038~\AA{} is more strongly absorbed by H$_2$ and possibly by \ion{C}{iv} lines. The observed FWHM is about 200~km~s$^{-1}$ (Fig.~\ref{sim_line_profile}).
Adding turbulent broadening to our 1-D simulation results in more symmetric line profiles, the red wing can be fitted with $v_{\mathrm{turb}}=150$~km~s$^{-1}$, but no blue wing is observed. In principle, this could be due to wind absorption. In the case of TW~Hya however, \citet{2007ApJ...655..345J} were able to demonstrate that a hot wind is incompatible with, for example, the observed \ion{C}{iv} line profiles.

\subsubsection{Model extensions}
Depending on the accretion geometry, radiation from the post-shock zone may be reprocessed in the pre-shock accretion flow. This will form a photoionised radiative precursor, sufficiently hot to emit in \ion{C}{iii} and even \ion{O}{vi} lines. The emitting region however is located close to the shock, where the material is moving at close to free-fall speed, and emission is redshifted by about 500~km~s$^{-1}$. \citet{2003ARep...47..498L} performed detailed 3-level-atom calculations supporting the above considerations with numerical simulations. He found that the total emission should be observed as two distinct peaks, one close to free-fall velocity about 500~km~s$^{-1}$ and the other formed post-shock. Only if the line of sight is perpendicular to the direction of flow these two peaks can coincide, but close to the rest wavelength. For TW~Hya, which is seen close to pole-on, this would require accretion in an equatorial layer. 

Figure~\ref{sim_line_profile} shows that in principle a large turbulent broadening could explain the observed line width, but it requires turbulent velocities larger than the post-shock bulk velocity. 
Nevertheless other plasma processes such as magnetic waves may supply the required turbulence.
 
The line widths could be explained if the emission originates
in the funnels well before entering the radiative precursor, in a region high above the stellar surface. A single feature corotating with the star would be observed moving through the line profile, as the star rotates, and a combination of many emission regions may result in a very broad line.
In this case an efficient and currently unknown heating mechanism in the accretion funnels is needed to form the hot ions.
The resolution of the current X-ray telescopes \emph{Chandra} and \emph{XMM-Newton} is insufficient to detect line broadening less than about 1000~km~s$^{-1}$ or line shifts less than a few 100~km~s$^{-1}$. We are therefore unable to compare our line-profile measurements with those for hotter ion lines, which are most likely accretion-dominated.

\subsection{Turbulence in boundary layers}

\label{turbulence}
Turbulence is an obvious way to generate significant line widths.
Turbulent flows convert bulk kinetic motion into heat and vorticity and a flow along a boundary layer would be an obvious
origin of turbulence. If, alternatively, accretion proceeds in an equatorial boundary layer, the difference in velocity is given by the difference between the stellar rotation $v_{\mathrm{rot}}$, and the velocity of material accreting close to the Keplerian velocity at the stellar surface $v_{\mathrm{Kepler}}=\sqrt{\frac{GM_*}{R_*}}\approx440$~km~s$^{-1}$, where $G$ is the gravitational constant, and the stellar parameters radius $R_*$ and mass $M_*$ for the example of TW~Hya are taken from \citet{1999ApJ...512L..63W}. The equatorial velocity can be estimated from $v_{\mathrm{eq}}=\frac{v\sin i}{\sin i}\approx40$~km~s$^{-1}$. 

In order to estimate the dissipated energy, we use the turbulent Crocco-Busemann relation, which is valid for turbulent flow along a surface. We follow the treatment by \citet{white} and neglect the initial temperature of the infalling flow, since the accretion disk is cool, and the photospheric temperature. We obtain a relation between temperature in the turbulent layer $T_{\mathrm{turb}}$ and velocity $u$ of flow: 
\begin{equation} 
T_{\mathrm{turb}}(u)=\frac{u}{2c_p}(\Delta v-u) \ ,\label{crocco}
\end{equation}
where $\Delta v=v_{\mathrm{Kepler}}-v_{\mathrm{eq}}$, $u$ is the velocity of the gas relative to the photosphere and $c_p$ has the usual thermodynamic meaning as specific heat at constant pressure. For an ideal gas, Eqn~(\ref{crocco}) simplifies to $T_{\mathrm{turb}}=2.3\times10^5\mathrm{ K}\frac{\Delta v^2}{\left(100\mathrm{ km s}^{-1}\right)^2}(\tilde{u}-\tilde{u}^2)$, with $\tilde{u}=\frac{u}{\Delta v}$, so the maximum temperature is reached where the infalling gas has slowed down by $\frac{\Delta v}{2}$. For $\Delta v \approx 400$~km~s$^{-1}$, the maximum temperature is $9\times10^5$~K and $2.3\times10^5$~K for half the Keplerian velocity. In the example of TW~Hya, in both scenarios the formation temperature of \ion{O}{vi} is reached. Because the line-of-sight is perpendicular to equatorial accretion, the peak of the velocities should
appear close to the rest wavelength with a line width comparable to the turbulent velocity. In this scenario, it is clearly
difficult to explain the shift of the line centre, but this process may  contribute to the emission responsible for a broad component, especially in stars with with only slightly shifted emission such as DF~Tau, V4046~Sgr and TWA~5 or in TW~Hya during the \emph{GHRS} and \emph{STIS} observations. 
For TWA 5, the scenario furthermore explains how accretion, as supported by a measured H$\alpha$ EW of 13.4~\AA, could proceed. This is in spite of the fact that the X-ray radiation originates in a low-density region, probably a corona, in contrast to that expected in a scenario of magnetically-funnelled accretion in hot spots.

\subsection{Red-shifted emission}

All UV observations of TW~Hya
show consistently red-shifted lines. The only possible explanation are matter flows onto the stellar surface because the far side of the star is blocked from view by the star itself. On the other hand, as argued above, the absence of 
a double-peaked profile with pre- and post-shock emission
implies that accretion spots are unlikely to explain their origin. We therefore now discuss a contribution from the accretion funnels.
Three-dimensional MHD simulations of \citet{2004ApJ...610..920R} give accretion spots with inhomogeneous velocity profiles, where the flow speed is close to free-fall at the spot centre and much slower at the edges. Their simulations  use ideal MHD equations and do not take into account the heating by turbulence caused by the difference in velocity within the accretion streams. In general the discussion in Sect.~\ref{turbulence} applies with velocity differences of the order of half the Keplerian velocity, so significant turbulent heating could arise in the funnels themselves, if turbulence is not suppressed by the strong magnetic fields. Further modelling and clarification of the accretion geometry is required to check the resulting emission measures.  Figure~\ref{sim_line_profile} shows that the shock, given some turbulent broadening in the post-shock zone, can account for a significant proportion of the observed emission in accordance with our post-shock cooling zone simulations.

Interestingly, in MS stars lines red-shifted by a few km~s$^{-1}$ are also observed \citep{1980ApJ...241..279A,1983ApJ...274..801A}, so there has to be an emission mechanism unrelated to accretion, usually attributed to downflows in the stellar transition region. This phenomenon may occur in CTTS as well.

\subsection{Other emission origins?}
Emission in the disk surrounding the star may in principle explain broad M-shaped lines, but the temperature at the disk surface is only a few thousand K  \citep{2004ApJ...607..369H}. Neither \ion{C}{iii} nor \ion{O}{vi} forms there. 

Also, for e.g. TW~Hya, 
where exceptionally broad lines are detected, the projected velocity of the disk at the inner disk edge is only $\approx20$~km~s$^{-1}$.  We observe no variation in the peak shift between the two observations of TW~Hya, so it seems unlikely that it is produced by a single feature moving with the rotating disk.

If the lines originate in an extended chromosphere, from the ratio of equatorial velocity and line width we can estimate the height of the emitting structures and find, using the values for \ion{O}{vi} 1032~\AA{} from Table~\ref{tab_gaussfits}, that the emission is placed at the stellar surface for T~Tau and at about three stellar radii for RU~Lup, V4046~Sgr, GM~Aur and TW~Hya. This can only explain line shifts if we assume upflowing material for RU~Lup and downflows for TW~Hya stable over all exposures, which seems unlikely.

\subsection{Variability} 
In theoretical simulations accretion, jets and wind generation are highly variable processes \citep{2006AN....327...53V}. 
Comparing the fluxes of TW Hya in the multiplet \ion{C}{iii} 1175~\AA{} , in observations acquired using \emph{FUSE} and \emph{HST}, we find that the flux decreases by approximately one quarter over three years. In contrast, a flux difference of  over a factor of five is measured for RU Lup observations \citep{2005AJ....129.2777H}.
 For TW~Hya, there are also \emph{FUSE} observations separated by several years. Unfortunately, the earlier observations are very short and of low SNR. Comparing the \emph{GHRS} and \emph{STIS} observations, separated by 4-8 years, the observed line width and shifts appear to be roughly consistent. 
We next considered correlations between these observations and FUSE data. Excluding the \ion{C}{iii} 1175~\AA{} multiplet, for which a line-profile analysis is impossible, there are no lines that are in common between the two sets of data. It is expected however, that lines of \ion{Si}{iv}, \ion{C}{iv} and \ion{N}{v} will originate in emission-line regions for which \ion{C}{iii} and \ion{O}{vi} emission is detected.
In all cases, the line widths fall in the range 100-300~km~s$^{-1}$, and for RU~Lup, T~Tau and DF~Tau the lines are all consistently blueshifted, but in the \emph{HST} observations to a much lesser extent than in the \emph{FUSE} data, conversely, the TW~Hya lines are more redshifted in the \emph{FUSE} data. This hints at a strong variability of CTTS, where the dominant UV emission mechanism may change between winds and accretion.

Longer monitoring is required to test if the difference between the objects in our sample is due to different phases of activity and accretion rate or if there are other fundamental differences.
This is similar to the situation in X-rays, where several luminosity changes are observed, most likely due to coronal flaring on the time scale of hours \citep{2002ApJ...567..434K,v4046}, but the evidence is inconclusive and some variability might also be due to time-variable accretion rates.

\section{Summary}
\label{summary} 
We have presented the hot ion lines originating from \ion{C}{iii} and \ion{O}{vi} in all available CTTS spectra observed with \emph{FUSE}. The fitted centroids range from about -170~km~s$^{-1}$ to +100~km~s$^{-1}$ and the shifts of the \ion{C}{iii} 977~\AA{} line and the \ion{O}{vi} 1032~\AA{} and 1038~\AA{} doublet are consistent. Most, if not all, lines are superrotationally broadened. The blue-shifted lines could originate in a stellar outflow, maybe a jet; the red-shifted lines are incompatible with current models of magnetospheric accretion. The presence of boundary layers, winds and extended chromospheres is inconsistent with the observed line profiles.

Furthermore, the absorbing column densities observed in X-rays are incompatible
with the optical reddening measured assuming a standard gas-to-dust ratios and  interstellar extinction law. This situation holds for the most heavily-veiled objects AA~Tau, DF~Tau, T~Tau, SU~Aur and, although for lower absolute values of reddening, RU~Lup.
In cases where data are available from both UV and X-ray, the corresponding objects demonstrate blue-shifted emission in the \emph{FUSE} observations.
We interpret this as a sign of a dust-free absorber, which may consist of shock-heated jet material.
In the ratio between the \ion{O}{vii} luminosity and the \ion{O}{vi} luminosity, all CTTS display a clear excess of soft X-ray emission in the \ion{O}{vii} lines. Together with X-ray results, this illustrates that the signatures of a hot accretion spot are most evident in the temperature range 1-2~MK.

\begin{acknowledgements}
% This work uses observations from XMM-Newton, an ESA science mission with instruments and contributions directly funded by ESA Member States and the USA (NASA) and from \emph{Chandra}.
We thank G.~Herczeg and C.~Johns-Krull, the referee, for their competent remarks, especially on the \emph{FUSE} wavelength calibration.
The data presented in this paper were obtained from the Multimission Archive at the Space Telescope Science Institute (MAST). STScI is operated by the Association of Universities for Research in Astronomy, Inc., under NASA contract NAS5-26555. Support for MAST for non-HST data is provided by the NASA Office of Space Science via grant NAG5-7584 and by other grants and contracts.\\
CHIANTI is a collaborative project involving the NRL (USA), RAL (UK), MSSL (UK), the Universities of Florence (Italy) and Cambridge (UK), and George Mason University (USA).\\
H.M.G. acknowledges support from DLR under 50OR0105.
\end{acknowledgements}
\bibliographystyle{aa} % style aa.bst
\bibliography{../articles}

\begin{thebibliography}{113}
\expandafter\ifx\csname natexlab\endcsname\relax\def\natexlab#1{#1}\fi

\bibitem[{{Ake} {et~al.}(2000){Ake}, {Dupree}, {Young}, {Linsky}, {Malina},
  {Griffiths}, {Siegmund}, \& {Woodgate}}]{2000ApJ...538L..87A}
{Ake}, T.~B., {Dupree}, A.~K., {Young}, P.~R., {et~al.} 2000, \apjl, 538, L87

\bibitem[{{Alencar} \& {Basri}(2000)}]{2000AJ....119.1881A}
{Alencar}, S.~H.~P. \& {Basri}, G. 2000, \aj, 119, 1881

\bibitem[{{Alencar} \& {Batalha}(2002)}]{2002ApJ...571..378A}
{Alencar}, S.~H.~P. \& {Batalha}, C. 2002, \apj, 571, 378

\bibitem[{{Ardila} {et~al.}(2002{\natexlab{a}}){Ardila}, {Basri}, {Walter},
  {Valenti}, \& {Johns-Krull}}]{2002ApJ...566.1100A}
{Ardila}, D.~R., {Basri}, G., {Walter}, F.~M., {Valenti}, J.~A., \&
  {Johns-Krull}, C.~M. 2002{\natexlab{a}}, \apj, 566, 1100

\bibitem[{{Ardila} {et~al.}(2002{\natexlab{b}}){Ardila}, {Basri}, {Walter},
  {Valenti}, \& {Johns-Krull}}]{2002ApJ...567.1013A}
{Ardila}, D.~R., {Basri}, G., {Walter}, F.~M., {Valenti}, J.~A., \&
  {Johns-Krull}, C.~M. 2002{\natexlab{b}}, \apj, 567, 1013

\bibitem[{{Argiroffi} {et~al.}(2007){Argiroffi}, {Maggio}, \&
  {Peres}}]{2007A&A...465L...5A}
{Argiroffi}, C., {Maggio}, A., \& {Peres}, G. 2007, \aap, 465, L5

\bibitem[{{Argiroffi} {et~al.}(2005){Argiroffi}, {Maggio}, {Peres}, {Stelzer},
  \& {Neuh{\"a}user}}]{twa5}
{Argiroffi}, C., {Maggio}, A., {Peres}, G., {Stelzer}, B., \& {Neuh{\"a}user},
  R. 2005, \aap, 439, 1149

\bibitem[{{Ayres}(2005)}]{2005ESASP.560..419A}
{Ayres}, T.~R. 2005, in ESA Special Publication, Vol. 560, 13th Cambridge
  Workshop on Cool Stars, Stellar Systems and the Sun, ed. F.~{Favata} \& {et
  al.}, 419--+

\bibitem[{{Ayres} \& {Linsky}(1980)}]{1980ApJ...241..279A}
{Ayres}, T.~R. \& {Linsky}, J.~L. 1980, \apj, 241, 279

\bibitem[{{Ayres} {et~al.}(1983){Ayres}, {Stencel}, {Linsky}, {Simon},
  {Jordan}, {Brown}, \& {Engvold}}]{1983ApJ...274..801A}
{Ayres}, T.~R., {Stencel}, R.~E., {Linsky}, J.~L., {et~al.} 1983, \apj, 274,
  801

\bibitem[{{Bally} {et~al.}(2007){Bally}, {Reipurth}, \&
  {Davis}}]{2007prpl.conf..215B}
{Bally}, J., {Reipurth}, B., \& {Davis}, C.~J. 2007, in Protostars and Planets
  V, ed. B.~{Reipurth}, D.~{Jewitt}, \& K.~{Keil}, 215--230

\bibitem[{{Balucinska-Church} \& {McCammon}(1992)}]{1992ApJ...400..699B}
{Balucinska-Church}, M. \& {McCammon}, D. 1992, \apj, 400, 699

\bibitem[{{Basri} \& {Batalha}(1990)}]{1990ApJ...363..654B}
{Basri}, G. \& {Batalha}, C. 1990, \apj, 363, 654

\bibitem[{{Bergin} {et~al.}(2004){Bergin}, {Calvet}, {Sitko}, {Abgrall},
  {D'Alessio}, {Herczeg}, {Roueff}, {Qi}, {Lynch}, {Russell}, {Brafford}, \&
  {Perry}}]{2004ApJ...614L.133B}
{Bergin}, E., {Calvet}, N., {Sitko}, M.~L., {et~al.} 2004, \apjl, 614, L133

\bibitem[{{Beristain} {et~al.}(2001){Beristain}, {Edwards}, \&
  {Kwan}}]{2001ApJ...551.1037B}
{Beristain}, G., {Edwards}, S., \& {Kwan}, J. 2001, \apj, 551, 1037

\bibitem[{{Bertout} \& {Genova}(2006)}]{2006A&A...460..499B}
{Bertout}, C. \& {Genova}, F. 2006, \aap, 460, 499

\bibitem[{{Bloemen}(1987)}]{1987ApJ...322..694B}
{Bloemen}, J.~B.~G.~M. 1987, \apj, 322, 694

\bibitem[{{Bouvier} {et~al.}(1986){Bouvier}, {Bertout}, {Benz}, \&
  {Mayor}}]{1986A&A...165..110B}
{Bouvier}, J., {Bertout}, C., {Benz}, W., \& {Mayor}, M. 1986, \aap, 165, 110

\bibitem[{{Bouvier} {et~al.}(1999){Bouvier}, {Chelli}, {Allain}, {Carrasco},
  {Costero}, {Cruz-Gonzalez}, {Dougados}, {Fern{\'a}ndez}, {Mart{\'{\i}}n},
  {M{\'e}nard}, {Mennessier}, {Mujica}, {Recillas}, {Salas}, {Schmidt}, \&
  {Wichmann}}]{1999A&A...349..619B}
{Bouvier}, J., {Chelli}, A., {Allain}, S., {et~al.} 1999, \aap, 349, 619

\bibitem[{{Calvet} \& {Gullbring}(1998)}]{calvetgullbring}
{Calvet}, N. \& {Gullbring}, E. 1998, \apj, 509, 802

\bibitem[{{Cardelli} {et~al.}(1989){Cardelli}, {Clayton}, \&
  {Mathis}}]{1989ApJ...345..245C}
{Cardelli}, J.~A., {Clayton}, G.~C., \& {Mathis}, J.~S. 1989, \apj, 345, 245

\bibitem[{{Chen} {et~al.}(1990){Chen}, {Simon}, {Longmore}, {Howell}, \&
  {Benson}}]{1990ApJ...357..224C}
{Chen}, W.~P., {Simon}, M., {Longmore}, A.~J., {Howell}, R.~R., \& {Benson},
  J.~A. 1990, \apj, 357, 224

\bibitem[{{Coffey} {et~al.}(2004){Coffey}, {Bacciotti}, {Woitas}, {Ray}, \&
  {Eisl{\"o}ffel}}]{2004ApJ...604..758C}
{Coffey}, D., {Bacciotti}, F., {Woitas}, J., {Ray}, T.~P., \& {Eisl{\"o}ffel},
  J. 2004, \apj, 604, 758

\bibitem[{{Cohen} \& {Kuhi}(1979)}]{1979ApJS...41..743C}
{Cohen}, M. \& {Kuhi}, L.~V. 1979, \apjs, 41, 743

\bibitem[{{Dere} {et~al.}(1998){Dere}, {Landi}, {Mason}, {Fossi}, \&
  {Young}}]{CHIANTII}
{Dere}, K.~P., {Landi}, E., {Mason}, H.~E., {Fossi}, B.~C.~M., \& {Young},
  P.~R. 1998, in ASP Conf. Ser. 143: The Scientific Impact of the Goddard High
  Resolution Spectrograph, 390--+

\bibitem[{{Dixon} {et~al.}(2007){Dixon}, {Sahnow}, {Barrett}, {Civeit},
  {Dupuis}, {Fullerton}, {Godard}, {Hsu}, {Kaiser}, {Kruk}, {Lacour},
  {Lindler}, {Massa}, {Robinson}, {Romelfanger}, \&
  {Sonnentrucker}}]{2007PASP..119..527D}
{Dixon}, W.~V., {Sahnow}, D.~J., {Barrett}, P.~E., {et~al.} 2007, \pasp, 119,
  527

\bibitem[{{Dupree} {et~al.}(2005{\natexlab{a}}){Dupree}, {Brickhouse}, {Smith},
  \& {Strader}}]{2005ApJ...625L.131D}
{Dupree}, A.~K., {Brickhouse}, N.~S., {Smith}, G.~H., \& {Strader}, J.
  2005{\natexlab{a}}, \apjl, 625, L131

\bibitem[{{Dupree} {et~al.}(2005{\natexlab{b}}){Dupree}, {Lobel}, {Young},
  {Ake}, {Linsky}, \& {Redfield}}]{2005ApJ...622..629D}
{Dupree}, A.~K., {Lobel}, A., {Young}, P.~R., {et~al.} 2005{\natexlab{b}},
  \apj, 622, 629

\bibitem[{{Edwards} {et~al.}(2006){Edwards}, {Fischer}, {Hillenbrand}, \&
  {Kwan}}]{2006ApJ...646..319E}
{Edwards}, S., {Fischer}, W., {Hillenbrand}, L., \& {Kwan}, J. 2006, \apj, 646,
  319

\bibitem[{{Eisl{\"o}ffel} \& {Mundt}(1998)}]{1998AJ....115.1554E}
{Eisl{\"o}ffel}, J. \& {Mundt}, R. 1998, \aj, 115, 1554

\bibitem[{{Eisner} {et~al.}(2006){Eisner}, {Chiang}, \&
  {Hillenbrand}}]{2006ApJ...637L.133E}
{Eisner}, J.~A., {Chiang}, E.~I., \& {Hillenbrand}, L.~A. 2006, \apjl, 637,
  L133

\bibitem[{{Errico} {et~al.}(2001){Errico}, {Lamzin}, \&
  {Vittone}}]{2001A&A...377..557E}
{Errico}, L., {Lamzin}, S.~A., \& {Vittone}, A.~A. 2001, \aap, 377, 557

\bibitem[{{Feldman} {et~al.}(2001){Feldman}, {Sahnow}, {Kruk}, {Murphy}, \&
  {Moos}}]{2001JGR...106.8119F}
{Feldman}, P.~D., {Sahnow}, D.~J., {Kruk}, J.~W., {Murphy}, E.~M., \& {Moos},
  H.~W. 2001, \jgr, 106, 8119

\bibitem[{{G{\'o}mez de Castro} \&
  {Ferro-Font{\'a}n}(2005)}]{2005MNRAS.362..569G}
{G{\'o}mez de Castro}, A.~I. \& {Ferro-Font{\'a}n}, C. 2005, \mnras, 362, 569

\bibitem[{{Gregory} {et~al.}(2006){Gregory}, {Jardine}, {Simpson}, \&
  {Donati}}]{2006MNRAS.371..999G}
{Gregory}, S.~G., {Jardine}, M., {Simpson}, I., \& {Donati}, J.-F. 2006,
  \mnras, 371, 999

\bibitem[{{G{\"u}del} {et~al.}(2005){G{\"u}del}, {Skinner}, {Briggs}, {Audard},
  {Arzner}, \& {Telleschi}}]{2005ApJ...626L..53G}
{G{\"u}del}, M., {Skinner}, S.~L., {Briggs}, K.~R., {et~al.} 2005, \apjl, 626,
  L53

\bibitem[{{G{\"u}del} {et~al.}(2007){G{\"u}del}, {Skinner}, {Mel'Nikov},
  {Audard}, {Telleschi}, \& {Briggs}}]{ttau}
{G{\"u}del}, M., {Skinner}, S.~L., {Mel'Nikov}, S.~Y., {et~al.} 2007, \aap,
  468, 529

\bibitem[{{G{\"u}del} \& {Telleschi}(2007)}]{manuelnh}
{G{\"u}del}, M. \& {Telleschi}, A. 2007, \aap, 474, L25

\bibitem[{{G{\"u}nther} {et~al.}(2006){G{\"u}nther}, {Liefke}, {Schmitt},
  {Robrade}, \& {Ness}}]{v4046}
{G{\"u}nther}, H.~M., {Liefke}, C., {Schmitt}, J.~H.~M.~M., {Robrade}, J., \&
  {Ness}, J.-U. 2006, \aap, 459, L29

\bibitem[{{G{\"u}nther} \& {Schmitt}(2007)}]{Bonn06}
{G{\"u}nther}, H.~M. \& {Schmitt}, J.~H.~M.~M. 2007, Memorie della Societa
  Astronomica Italiana, 78, 359

\bibitem[{{G{\"u}nther} {et~al.}(2007){G{\"u}nther}, {Schmitt}, {Robrade}, \&
  {Liefke}}]{acc_model}
{G{\"u}nther}, H.~M., {Schmitt}, J.~H.~M.~M., {Robrade}, J., \& {Liefke}, C.
  2007, \aap, 466, 1111

\bibitem[{{Hartigan} {et~al.}(1987){Hartigan}, {Raymond}, \&
  {Hartmann}}]{1987ApJ...316..323H}
{Hartigan}, P., {Raymond}, J., \& {Hartmann}, L. 1987, \apj, 316, 323

\bibitem[{{Hartmann} \& {Stauffer}(1989)}]{1989AJ.....97..873H}
{Hartmann}, L. \& {Stauffer}, J.~R. 1989, \aj, 97, 873

\bibitem[{{Herczeg} {et~al.}(2002){Herczeg}, {Linsky}, {Valenti},
  {Johns-Krull}, \& {Wood}}]{2002ApJ...572..310H}
{Herczeg}, G.~J., {Linsky}, J.~L., {Valenti}, J.~A., {Johns-Krull}, C.~M., \&
  {Wood}, B.~E. 2002, \apj, 572, 310

\bibitem[{{Herczeg} {et~al.}(2006){Herczeg}, {Linsky}, {Walter}, {Gahm}, \&
  {Johns-Krull}}]{2006ApJS..165..256H}
{Herczeg}, G.~J., {Linsky}, J.~L., {Walter}, F.~M., {Gahm}, G.~F., \&
  {Johns-Krull}, C.~M. 2006, \apjs, 165, 256

\bibitem[{{Herczeg} {et~al.}(2005){Herczeg}, {Walter}, {Linsky}, {Gahm},
  {Ardila}, {Brown}, {Johns-Krull}, {Simon}, \&
  {Valenti}}]{2005AJ....129.2777H}
{Herczeg}, G.~J., {Walter}, F.~M., {Linsky}, J.~L., {et~al.} 2005, \aj, 129,
  2777

\bibitem[{{Herczeg} {et~al.}(2004){Herczeg}, {Wood}, {Linsky}, {Valenti}, \&
  {Johns-Krull}}]{2004ApJ...607..369H}
{Herczeg}, G.~J., {Wood}, B.~E., {Linsky}, J.~L., {Valenti}, J.~A., \&
  {Johns-Krull}, C.~M. 2004, \apj, 607, 369

\bibitem[{{Hueso} \& {Guillot}(2005)}]{2005A&A...442..703H}
{Hueso}, R. \& {Guillot}, T. 2005, \aap, 442, 703

\bibitem[{{Johns-Krull} \& {Herczeg}(2007)}]{2007ApJ...655..345J}
{Johns-Krull}, C.~M. \& {Herczeg}, G.~J. 2007, \apj, 655, 345

\bibitem[{{Joy}(1949)}]{1949ApJ...110..424J}
{Joy}, A.~H. 1949, \apj, 110, 424

\bibitem[{{Kastner} {et~al.}(2002){Kastner}, {Huenemoerder}, {Schulz},
  {Canizares}, \& {Weintraub}}]{2002ApJ...567..434K}
{Kastner}, J.~H., {Huenemoerder}, D.~P., {Schulz}, N.~S., {Canizares}, C.~R.,
  \& {Weintraub}, D.~A. 2002, \apj, 567, 434

\bibitem[{{Kenyon} \& {Hartmann}(1995)}]{1995ApJS..101..117K}
{Kenyon}, S.~J. \& {Hartmann}, L. 1995, \apjs, 101, 117

\bibitem[{{Koenigl}(1991)}]{1991ApJ...370L..39K}
{Koenigl}, A. 1991, \apjl, 370, L39

\bibitem[{{Koresko}(2000)}]{2000ApJ...531L.147K}
{Koresko}, C.~D. 2000, \apjl, 531, L147

\bibitem[{{Lamzin}(1998)}]{lamzin}
{Lamzin}, S.~A. 1998, Astronomy Reports, 42, 322

\bibitem[{{Lamzin}(2003)}]{2003ARep...47..498L}
{Lamzin}, S.~A. 2003, Astronomy Reports, 47, 498

\bibitem[{{Lamzin} {et~al.}(1996){Lamzin}, {Bisnovatyi-Kogan}, {Errico},
  {Giovannelli}, {Katysheva}, {Rossi}, \& {Vittone}}]{1996A&A...306..877L}
{Lamzin}, S.~A., {Bisnovatyi-Kogan}, G.~S., {Errico}, L., {et~al.} 1996, \aap,
  306, 877

\bibitem[{{Lamzin} {et~al.}(2004){Lamzin}, {Kravtsova}, {Romanova}, \&
  {Batalha}}]{2004AstL...30..413L}
{Lamzin}, S.~A., {Kravtsova}, A.~S., {Romanova}, M.~M., \& {Batalha}, C. 2004,
  Astronomy Letters, 30, 413

\bibitem[{{Lamzin} {et~al.}(2001){Lamzin}, {Vittone}, \&
  {Errico}}]{2001AstL...27..313L}
{Lamzin}, S.~A., {Vittone}, A.~A., \& {Errico}, L. 2001, Astronomy Letters, 27,
  313

\bibitem[{{Landi} {et~al.}(2006){Landi}, {Del Zanna}, {Young}, {Dere}, {Mason},
  \& {Landini}}]{CHIANTIVII}
{Landi}, E., {Del Zanna}, G., {Young}, P.~R., {et~al.} 2006, \apjs, 162, 261

\bibitem[{{Lavalley-Fouquet} {et~al.}(2000){Lavalley-Fouquet}, {Cabrit}, \&
  {Dougados}}]{2000A&A...356L..41L}
{Lavalley-Fouquet}, C., {Cabrit}, S., \& {Dougados}, C. 2000, \aap, 356, L41

\bibitem[{{Mamajek} {et~al.}(2002){Mamajek}, {Meyer}, \&
  {Liebert}}]{2002AJ....124.1670M}
{Mamajek}, E.~E., {Meyer}, M.~R., \& {Liebert}, J. 2002, \aj, 124, 1670

\bibitem[{{Matt} {et~al.}(2002){Matt}, {Goodson}, {Winglee}, \&
  {B{\"o}hm}}]{2002ApJ...574..232M}
{Matt}, S., {Goodson}, A.~P., {Winglee}, R.~M., \& {B{\"o}hm}, K.-H. 2002,
  \apj, 574, 232

\bibitem[{{Mazzotta} {et~al.}(1998){Mazzotta}, {Mazzitelli}, {Colafrancesco},
  \& {Vittorio}}]{mazzottaetal}
{Mazzotta}, P., {Mazzitelli}, G., {Colafrancesco}, S., \& {Vittorio}, N. 1998,
  \aaps, 133, 403

\bibitem[{{McCandliss}(2003)}]{2003PASP..115..651M}
{McCandliss}, S.~R. 2003, \pasp, 115, 651

\bibitem[{{Metchev} {et~al.}(2004){Metchev}, {Hillenbrand}, \&
  {Meyer}}]{2004ApJ...600..435M}
{Metchev}, S.~A., {Hillenbrand}, L.~A., \& {Meyer}, M.~R. 2004, \apj, 600, 435

\bibitem[{{Mohanty} {et~al.}(2003){Mohanty}, {Jayawardhana}, \& {Barrado y
  Navascu{\'e}s}}]{2003ApJ...593L.109M}
{Mohanty}, S., {Jayawardhana}, R., \& {Barrado y Navascu{\'e}s}, D. 2003,
  \apjl, 593, L109

\bibitem[{{Muzerolle} {et~al.}(2000){Muzerolle}, {Calvet}, {Brice{\~ n}o},
  {Hartmann}, \& {Hillenbrand}}]{2000ApJ...535L..47M}
{Muzerolle}, J., {Calvet}, N., {Brice{\~ n}o}, C., {Hartmann}, L., \&
  {Hillenbrand}, L. 2000, \apjl, 535, L47

\bibitem[{{Muzerolle} {et~al.}(1998){Muzerolle}, {Hartmann}, \&
  {Calvet}}]{1998AJ....116.2965M}
{Muzerolle}, J., {Hartmann}, L., \& {Calvet}, N. 1998, \aj, 116, 2965

\bibitem[{{Najita} {et~al.}(2007){Najita}, {Strom}, \&
  {Muzerolle}}]{2007MNRAS.378..369N}
{Najita}, J.~R., {Strom}, S.~E., \& {Muzerolle}, J. 2007, \mnras, 378, 369

\bibitem[{{Ness} {et~al.}(2004){Ness}, {G{\"u}del}, {Schmitt}, {Audard}, \&
  {Telleschi}}]{2004A&A...427..667N}
{Ness}, J.-U., {G{\"u}del}, M., {Schmitt}, J.~H.~M.~M., {Audard}, M., \&
  {Telleschi}, A. 2004, \aap, 427, 667

\bibitem[{{Pudritz} {et~al.}(2007){Pudritz}, {Ouyed}, {Fendt}, \&
  {Brandenburg}}]{2007prpl.conf..277P}
{Pudritz}, R.~E., {Ouyed}, R., {Fendt}, C., \& {Brandenburg}, A. 2007, in
  Protostars and Planets V, ed. B.~{Reipurth}, D.~{Jewitt}, \& K.~{Keil},
  277--294

\bibitem[{{Qi} {et~al.}(2004){Qi}, {Ho}, {Wilner}, {Takakuwa}, {Hirano},
  {Ohashi}, {Bourke}, {Zhang}, {Blake}, {Hogerheijde}, {Saito}, {Choi}, \&
  {Yang}}]{2004ApJ...616L..11Q}
{Qi}, C., {Ho}, P.~T.~P., {Wilner}, D.~J., {et~al.} 2004, \apjl, 616, L11

\bibitem[{{Quast} {et~al.}(2000){Quast}, {Torres}, {de La Reza}, {da Silva}, \&
  {Mayor}}]{2000IAUS..200P..28Q}
{Quast}, G.~R., {Torres}, C.~A.~O., {de La Reza}, R., {da Silva}, L., \&
  {Mayor}, M. 2000, in IAU Symposium, 28P--+

\bibitem[{{Redfield} {et~al.}(2002){Redfield}, {Linsky}, {Ake}, {Ayres},
  {Dupree}, {Robinson}, {Wood}, \& {Young}}]{2002ApJ...581..626R}
{Redfield}, S., {Linsky}, J.~L., {Ake}, T.~B., {et~al.} 2002, \apj, 581, 626

\bibitem[{{Robrade} \& {Schmitt}(2006)}]{Rob0507}
{Robrade}, J. \& {Schmitt}, J.~H.~M.~M. 2006, \aap, 449, 737

\bibitem[{{Robrade} \& {Schmitt}(2007)}]{RULup}
{Robrade}, J. \& {Schmitt}, J.~H.~M.~M. 2007, \aap, 473, 229

\bibitem[{{Rodmann} {et~al.}(2006){Rodmann}, {Henning}, {Chandler}, {Mundy}, \&
  {Wilner}}]{2006A&A...446..211R}
{Rodmann}, J., {Henning}, T., {Chandler}, C.~J., {Mundy}, L.~G., \& {Wilner},
  D.~J. 2006, \aap, 446, 211

\bibitem[{{Rodriguez}(1995)}]{1995RMxAC...1....1R}
{Rodriguez}, L.~F. 1995, in Revista Mexicana de Astronomia y Astrofisica
  Conference Series, ed. S.~{Lizano} \& J.~M. {Torrelles}, 1--+

\bibitem[{{Romanova} {et~al.}(2004){Romanova}, {Ustyugova}, {Koldoba}, \&
  {Lovelace}}]{2004ApJ...610..920R}
{Romanova}, M.~M., {Ustyugova}, G.~V., {Koldoba}, A.~V., \& {Lovelace},
  R.~V.~E. 2004, \apj, 610, 920

\bibitem[{{Saar} \& {Osten}(1997)}]{1997MNRAS.284..803S}
{Saar}, S.~H. \& {Osten}, R.~A. 1997, \mnras, 284, 803

\bibitem[{{Sallmen} {et~al.}(2000){Sallmen}, {Johns-Krull}, {Welsh}, \&
  {Griffiths}}]{2000AAS...197.4708S}
{Sallmen}, S., {Johns-Krull}, C.~M., {Welsh}, B., \& {Griffiths}, N. 2000, in
  Bulletin of the American Astronomical Society, 1482--+

\bibitem[{{Salyk} {et~al.}(2007){Salyk}, {Blake}, {Boogert}, \&
  {Brown}}]{2007ApJ...655L.105S}
{Salyk}, C., {Blake}, G.~A., {Boogert}, A.~C.~A., \& {Brown}, J.~M. 2007,
  \apjl, 655, L105

\bibitem[{{Savage} \& {Mathis}(1979)}]{1979ARA&A..17...73S}
{Savage}, B.~D. \& {Mathis}, J.~S. 1979, \araa, 17, 73

\bibitem[{{Schaefer} {et~al.}(2006){Schaefer}, {Simon}, {Beck}, {Nelan}, \&
  {Prato}}]{2006AJ....132.2618S}
{Schaefer}, G.~H., {Simon}, M., {Beck}, T.~L., {Nelan}, E., \& {Prato}, L.
  2006, \aj, 132, 2618

\bibitem[{{Schegerer} {et~al.}(2006){Schegerer}, {Wolf}, {Voshchinnikov},
  {Przygodda}, \& {Kessler-Silacci}}]{2006A&A...456..535S}
{Schegerer}, A., {Wolf}, S., {Voshchinnikov}, N.~V., {Przygodda}, F., \&
  {Kessler-Silacci}, J.~E. 2006, \aap, 456, 535

\bibitem[{{Schmitt} \& {Robrade}(2007)}]{2007A&A...462L..41S}
{Schmitt}, J.~H.~M.~M. \& {Robrade}, J. 2007, \aap, 462, L41

\bibitem[{{Shu} {et~al.}(1994){Shu}, {Najita}, {Ostriker}, {Wilkin}, {Ruden},
  \& {Lizano}}]{1994ApJ...429..781S}
{Shu}, F., {Najita}, J., {Ostriker}, E., {et~al.} 1994, \apj, 429, 781

\bibitem[{{Simon} {et~al.}(2000){Simon}, {Dutrey}, \&
  {Guilloteau}}]{2000ApJ...545.1034S}
{Simon}, M., {Dutrey}, A., \& {Guilloteau}, S. 2000, \apj, 545, 1034

\bibitem[{{Stelzer} \& {Schmitt}(2004)}]{twhya}
{Stelzer}, B. \& {Schmitt}, J.~H.~M.~M. 2004, \aap, 418, 687

\bibitem[{{Stempels} \& {Gahm}(2004)}]{2004A&A...421.1159S}
{Stempels}, H.~C. \& {Gahm}, G.~F. 2004, \aap, 421, 1159

\bibitem[{{Stempels} {et~al.}(2007){Stempels}, {Gahm}, \&
  {Petrov}}]{2007A&A...461..253S}
{Stempels}, H.~C., {Gahm}, G.~F., \& {Petrov}, P.~P. 2007, \aap, 461, 253

\bibitem[{{Stempels} \& {Piskunov}(2002)}]{2002A&A...391..595S}
{Stempels}, H.~C. \& {Piskunov}, N. 2002, \aap, 391, 595

\bibitem[{{Strom} {et~al.}(1989){Strom}, {Strom}, {Edwards}, {Cabrit}, \&
  {Skrutskie}}]{1989AJ.....97.1451S}
{Strom}, K.~M., {Strom}, S.~E., {Edwards}, S., {Cabrit}, S., \& {Skrutskie},
  M.~F. 1989, \aj, 97, 1451

\bibitem[{{Takami} {et~al.}(2001){Takami}, {Bailey}, {Gledhill},
  {Chrysostomou}, \& {Hough}}]{2001MNRAS.323..177T}
{Takami}, M., {Bailey}, J., {Gledhill}, T.~M., {Chrysostomou}, A., \& {Hough},
  J.~H. 2001, \mnras, 323, 177

\bibitem[{{Telleschi} {et~al.}(2007){Telleschi}, {G{\"u}del}, {Briggs},
  {Skinner}, {Audard}, \& {Franciosini}}]{ABAur}
{Telleschi}, A., {G{\"u}del}, M., {Briggs}, K.~R., {et~al.} 2007, \aap, 468,
  541

\bibitem[{{Torres} {et~al.}(2003){Torres}, {Guenther}, {Marschall},
  {Neuh{\"a}user}, {Latham}, \& {Stefanik}}]{2003AJ....125..825T}
{Torres}, G., {Guenther}, E.~W., {Marschall}, L.~A., {et~al.} 2003, \aj, 125,
  825

\bibitem[{{Uchida} {et~al.}(2004){Uchida}, {Calvet}, {Hartmann}, {Kemper},
  {Forrest}, {Watson}, {D'Alessio}, {Chen}, {Furlan}, {Sargent}, {Brandl},
  {Herter}, {Morris}, {Myers}, {Najita}, {Sloan}, {Barry}, {Green}, {Keller},
  \& {Hall}}]{2004ApJS..154..439U}
{Uchida}, K.~I., {Calvet}, N., {Hartmann}, L., {et~al.} 2004, \apjs, 154, 439

\bibitem[{{Uchida} \& {Shibata}(1984)}]{1984PASJ...36..105U}
{Uchida}, Y. \& {Shibata}, K. 1984, \pasj, 36, 105

\bibitem[{{Unruh} {et~al.}(1998){Unruh}, {Collier Cameron}, \&
  {Guenther}}]{1998MNRAS.295..781U}
{Unruh}, Y.~C., {Collier Cameron}, A., \& {Guenther}, E. 1998, \mnras, 295, 781

\bibitem[{{Valenti} {et~al.}(2000){Valenti}, {Johns-Krull}, \&
  {Linsky}}]{2000ApJS..129..399V}
{Valenti}, J.~A., {Johns-Krull}, C.~M., \& {Linsky}, J.~L. 2000, \apjs, 129,
  399

\bibitem[{{van Langevelde} {et~al.}(1994{\natexlab{a}}){van Langevelde}, {van
  Dishoeck}, \& {Blake}}]{1994ApJ...425L..45V}
{van Langevelde}, H.~J., {van Dishoeck}, E.~F., \& {Blake}, G.~A.
  1994{\natexlab{a}}, \apjl, 425, L45

\bibitem[{{van Langevelde} {et~al.}(1994{\natexlab{b}}){van Langevelde}, {van
  Dishoeck}, {van der Werf}, \& {Blake}}]{1994A&A...287L..25V}
{van Langevelde}, H.~J., {van Dishoeck}, E.~F., {van der Werf}, P.~P., \&
  {Blake}, G.~A. 1994{\natexlab{b}}, \aap, 287, L25

\bibitem[{{Vilhu} {et~al.}(1987){Vilhu}, {Gustafsson}, \&
  {Edvardsson}}]{1987ApJ...320..850V}
{Vilhu}, O., {Gustafsson}, B., \& {Edvardsson}, B. 1987, \apj, 320, 850

\bibitem[{{von Rekowski} \& {Brandenburg}(2006)}]{2006AN....327...53V}
{von Rekowski}, B. \& {Brandenburg}, A. 2006, Astronomische Nachrichten, 327,
  53

\bibitem[{{Vuong} {et~al.}(2003){Vuong}, {Montmerle}, {Grosso}, {Feigelson},
  {Verstraete}, \& {Ozawa}}]{2003A&A...408..581V}
{Vuong}, M.~H., {Montmerle}, T., {Grosso}, N., {et~al.} 2003, \aap, 408, 581

\bibitem[{{Walter} {et~al.}(2003){Walter}, {Herczeg}, {Brown}, {Ardila},
  {Gahm}, {Johns-Krull}, {Lissauer}, {Simon}, \&
  {Valenti}}]{2003AJ....126.3076W}
{Walter}, F.~M., {Herczeg}, G., {Brown}, A., {et~al.} 2003, \aj, 126, 3076

\bibitem[{{Webb} {et~al.}(1999){Webb}, {Zuckerman}, {Platais}, {Patience},
  {White}, {Schwartz}, \& {McCarthy}}]{1999ApJ...512L..63W}
{Webb}, R.~A., {Zuckerman}, B., {Platais}, I., {et~al.} 1999, \apjl, 512, L63

\bibitem[{{Weinberger} {et~al.}(2004){Weinberger}, {Becklin}, {Zuckerman}, \&
  {Song}}]{2004AJ....127.2246W}
{Weinberger}, A.~J., {Becklin}, E.~E., {Zuckerman}, B., \& {Song}, I. 2004,
  \aj, 127, 2246

\bibitem[{{White}(1991)}]{white}
{White}, F.~M. 1991, {Viscous fluid flow} (New York: McGraw-Hill, 1991)

\bibitem[{{Wilkinson} {et~al.}(2002){Wilkinson}, {Harper}, {Brown}, \&
  {Herczeg}}]{2002AJ....124.1077W}
{Wilkinson}, E., {Harper}, G.~M., {Brown}, A., \& {Herczeg}, G.~J. 2002, \aj,
  124, 1077

\bibitem[{{Wood} {et~al.}(1997){Wood}, {Linsky}, \&
  {Ayres}}]{1997ApJ...478..745W}
{Wood}, B.~E., {Linsky}, J.~L., \& {Ayres}, T.~R. 1997, \apj, 478, 745

\bibitem[{{Zel'Dovich} \& {Raizer}(1967)}]{raizerzeldovich}
{Zel'Dovich}, Y.~B. \& {Raizer}, Y.~P. 1967, {Physics of shock waves and
  high-temperature hydrodynamic phenomena} (New York: Academic Press,
  1966/1967, edited by Hayes, W.D.; Probstein, Ronald F.)

\end{thebibliography}
\end{document}